# In Memory of Prof. Hidekuni Takekoshi

10/26/1926–1/11/2020

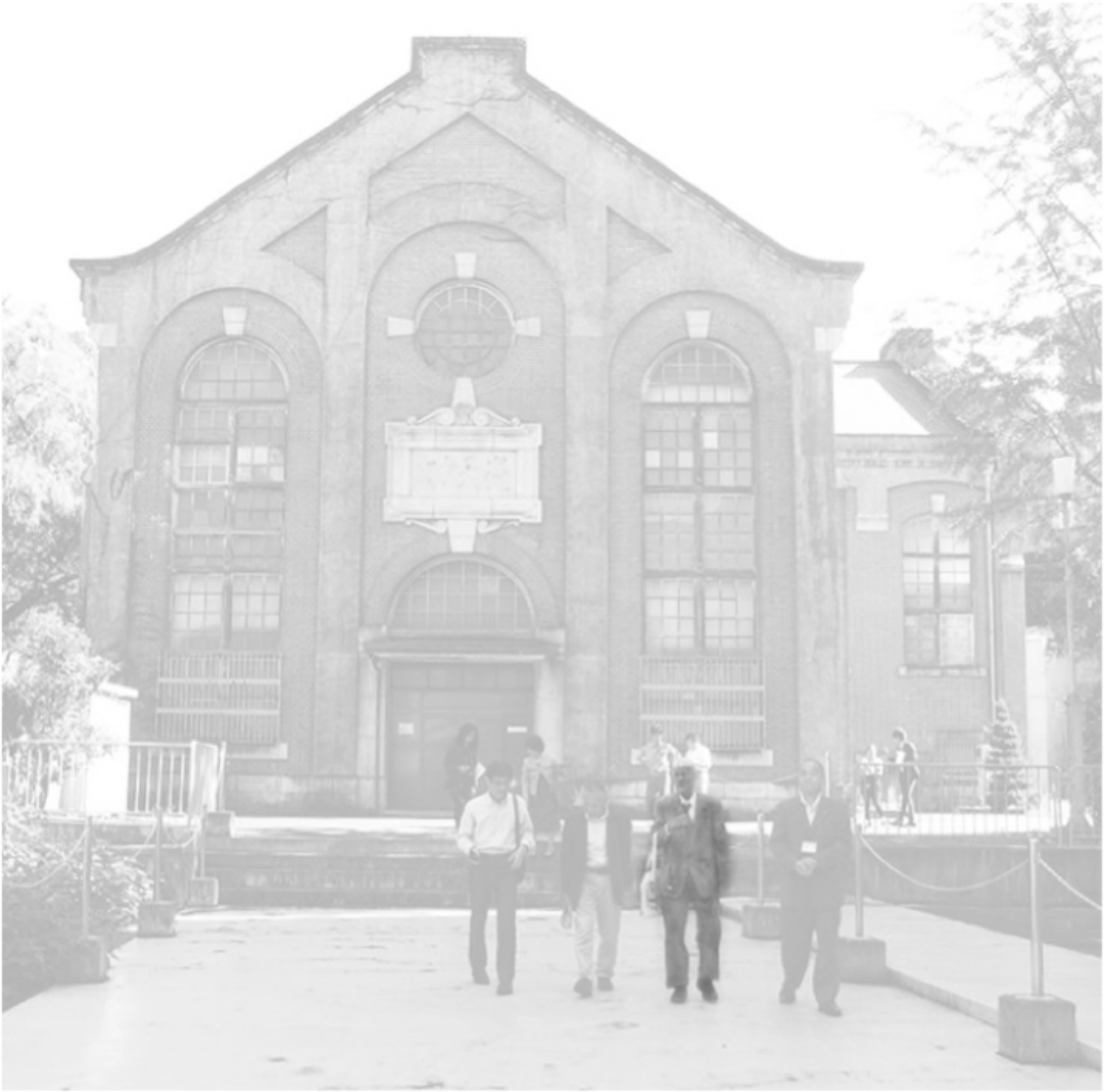

# In Memory of Prof. Hidekuni Takekoshi

10/26/1926–1/11/2020

**A memorial to Prof. Hidekuni Hidekoshi, Kyoto University, Japan – an accelerator pioneer in Japan, teacher, mentor, friend, unassuming but knowing his accomplishments and worth, frugal but generous, enjoying life. Japanese contributions given in Japanese and English, English in English.**

**Contents**

## Professional Profile

竹腰秀邦先生を偲んで

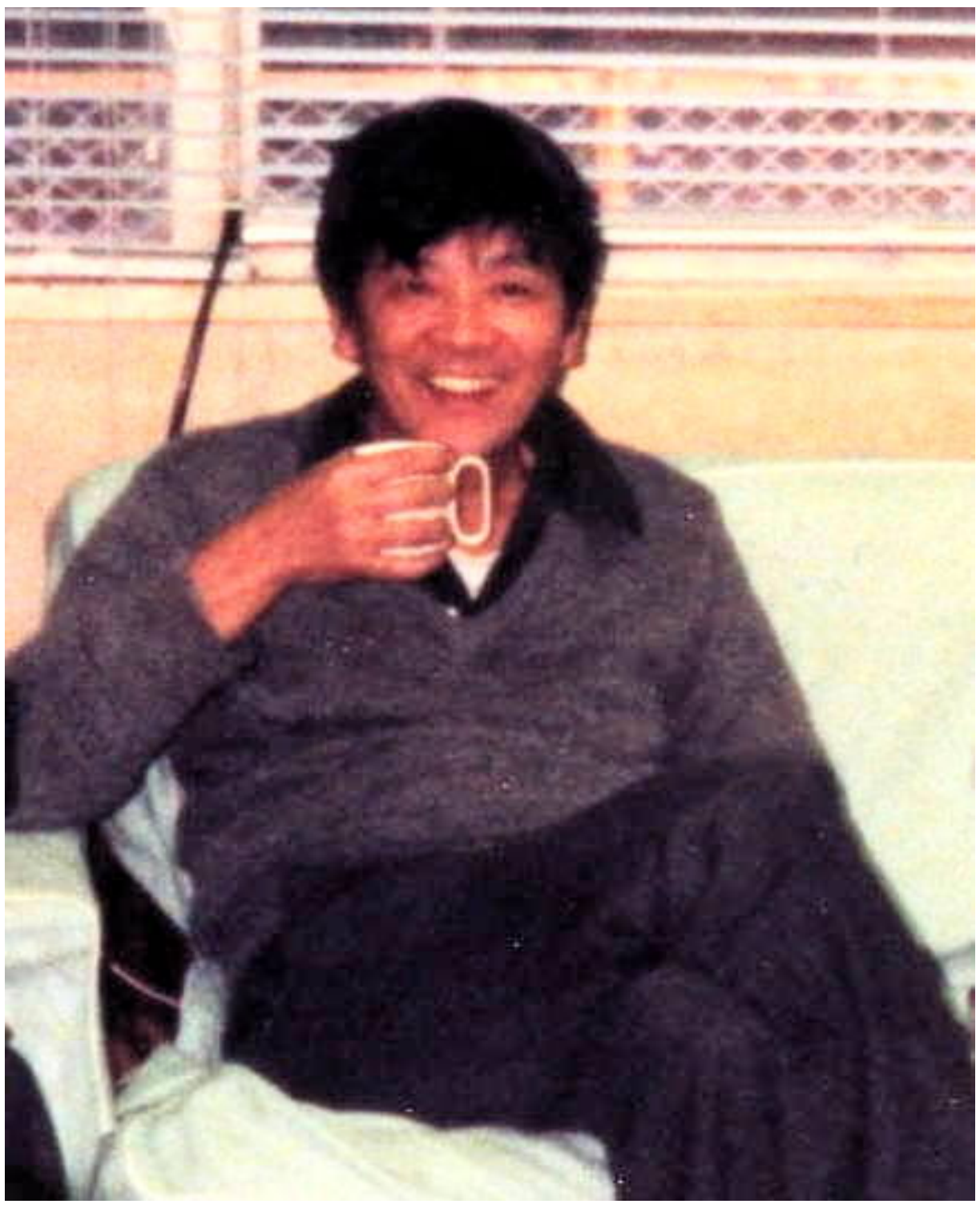

図１：1985年原研にて

At JAERI in 1985

誠に残念な事でありますが，元原子力研究所主任研究員、元京都大学化学研究所教授、　竹腰秀邦　京都大学名誉教授におかれましては，本年1月11日，お亡くなりになりました．享年93歳でした．　彼は台湾で生まれ育ちました。

竹腰秀邦先生は，敗戦当時京都大学荒勝研究室の学生で、戦後旧荒勝研究室の一員としてサイクロトロンの復興に参加し、1955年に組み立てを完了されました。その年の11月にはビームを外部に取り出し、放射性同位元素の製造が開始されました。建屋は琵琶湖疎水を活用して日本初の商用電力発電所となった、当時築９０年の蹴上発電所の二代目の建物で、既に廃屋となって雨漏りがする状態だった建屋を京都市から借りたようです。

その後1956年から旧日本原子力研究所（原研、現日本原子力研究開発機構）の物理部に所属されました。原研では、日本でいち早く原子炉からの熱中性子を用いてメスバウアー効果の研究を開始されました。また、1962年からは電子リニアックの室長として加速器の開発・運転・維持および電

子リニアックからの制動輻射線による中性子を用いた中性子核物理等の研究を指導されました。当初の原研電子リニアックは加速エネルギー20MeVで、当時国内では最新鋭の性能を誇るものでしたが、さらなる高性能化の要請を受けて、1969年から約3年かけて加速エネルギー120MeVへの増力が行われました（JAERI レポート 1238, 1975）。この増力には極めて限られた予算のなかで、竹腰室長のもと室員17名が加速器機器の製作を分担し、加速管及びクライストロンなどを除くほぼ全てを自主製作するというものでした。以降この加速器を用いた大電力バンチャーの開発、導波管用マイクロ波窓の開発、大型クライストロン用Ba含侵カソードの開発などの多くの研究開発が実施されました。また、実験装置の製作においても190mの中性子flight path等の建設などを始めとして様々な測定装置が設置されました。測定器の製作に当たっては竹腰室長自らハンドメイド装置の製作に当たられました。電子リニアックが“竹腰色満載の”所謂手作りであったことが、その後の絶え間ない改良と改修により、ビーム出力と質の向上を可能とした一因であったといえます。その結果原研電子リニアックは

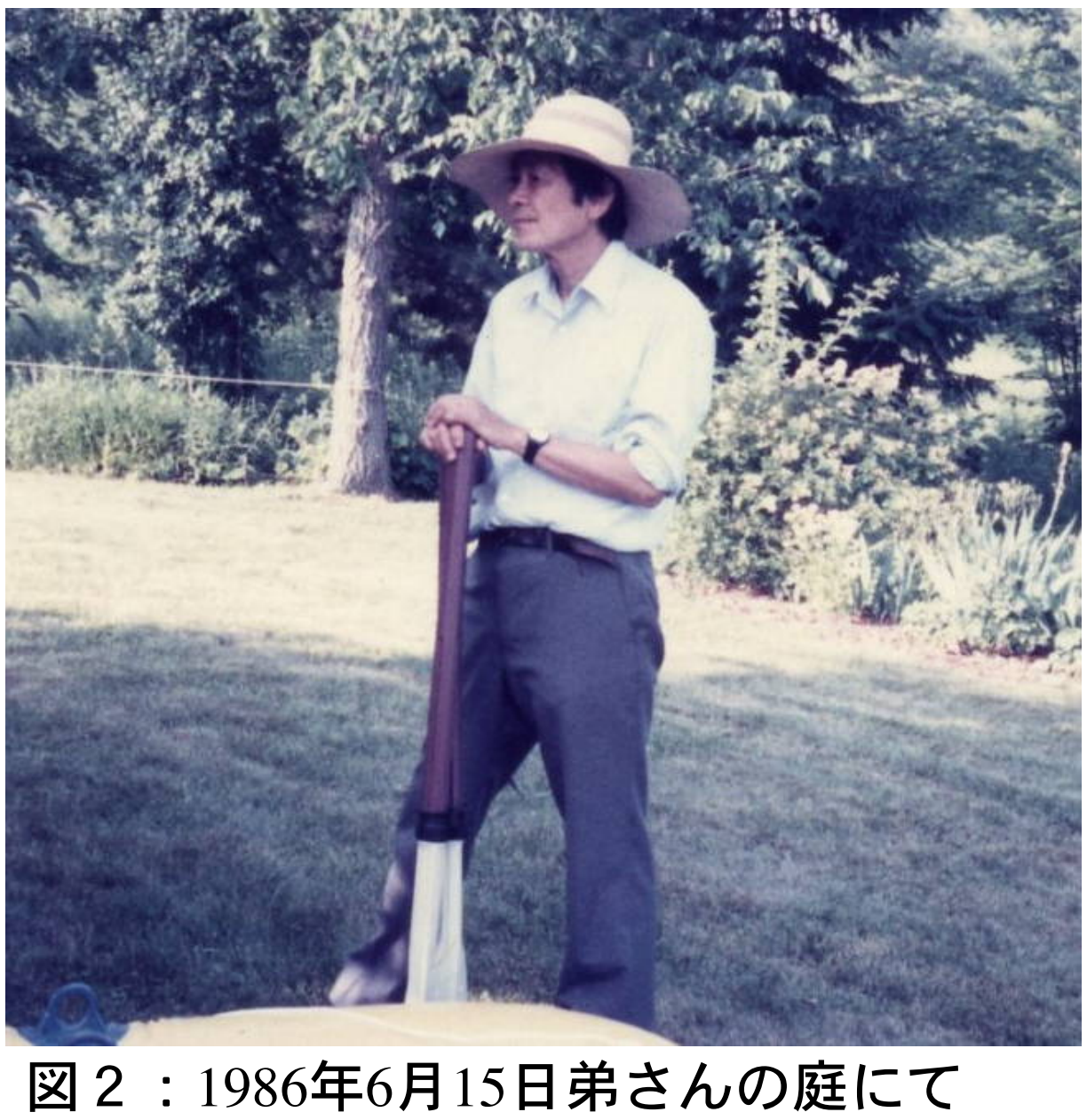

図２：1986年6月15日弟さんの庭にて
June 15, 1986, in my brother's garden

1993年までの21年におよぶ運転の継続と多様な研究にビームを供することができました。個人的には、竹腰室長は“竹腰のおじさん”（竹腰夫人も同じ物理部に所属しておられたため）と呼ばれて若い研究者への面倒見の良さ、技術サポートスタップからの信頼の厚さには定評がありました。

1976年からは京都大学化学研究所　原子核科学研究施設（旧蹴上分室）に赴任されました。京都大学化学研究所教授として戦後立ち上げに尽力されたサイクロトロンを

支えに戻ってこられたことになります。当時、先生には理学研究科からだけでなく、原子核工学や他大学の学生も委託で来ていて、彼らと共に、サイクロトロンのビームを使った生物・材料照射実験や、サイクロトロンビームの引出効率向上のため、引出部に magnetic channel を付加するなどの改造もされました。手作りの特技から自作したスメアー試料交換回転台（検出器はアロカから買ったＧＭ管）の回路・制御を、岩下がＰＣ８８０１等を用いて構成させてもらいました。アーク溶接等もご自分でされていました。夏の暑い時期にランニングシャツだけで施工し、防護マスクはされていましたが、発生する紫外線でかなり日焼け？されていた。これによりアーク溶接に関する知見を深めさせてもらいました。

一方、超伝導サイクロトロンの研究も進められ、磁石や共振器の試作もされました。また、京都大学の第３キャンパス構想が京阪奈地区を想定して練られていた頃、それに合流すべく、800MeV陽子線形加速器と、それを入射器とする30GeVシンクロトロンで構成される、共用加速器計画等を学内の関係者と共に進められました。バブル崩壊と共に立ち消えになりましたが、これはある意味、国内で継承されているような気がします。

前述のように建屋は旧蹴上発電所のものだったため、地下には排水路の名残があり、メインの水路は閉じられてサイクロトロンの冷却水となる８０トンを擁するプールとなっていました。閉じられた下流側に小さな池が出来ていて、ボウフラ対策のためか、そこに鯉を放ち、時々パンの耳を餌として与えていらっしゃいました。

1985年頃からは、老朽化したサイクロを廃止し、蹴上地区から宇治地区へ移転してイオン線形加速器を建設するという計画に井上信先生とご尽力されました。ご退職の当日1990年3月31日に返還書類を渡せたとのことで、蹴上サイクロトロンの誕生と終結に付き合われたことになります。（「加速器」Vol.3, No.4, pp.384-390,(2006)、Vol.4, No.1, pp.18-23, (2007)、化学研究所「黄檗」No.29, 13-1 ）蹴上の建物撤収時に先の鯉を回収？する際、鯉が釣られてしまったとぼやかれていたようです。その鯉は宇治地区にあった池に放たれたと聞いていますが、その池も今はありません。蹴上の建屋は現在関西電力が管理していて、時々一般公開されているようです。

ちょうど移転計画が動き出す頃、親交があったロスアラモス研究所の R. A. Jameson 博士が滞在されていました。

Keageでは5人の博士号を含むおよそ16人の学生が訓練を受けました。

ご定年後、一年をおいて広島電機大学

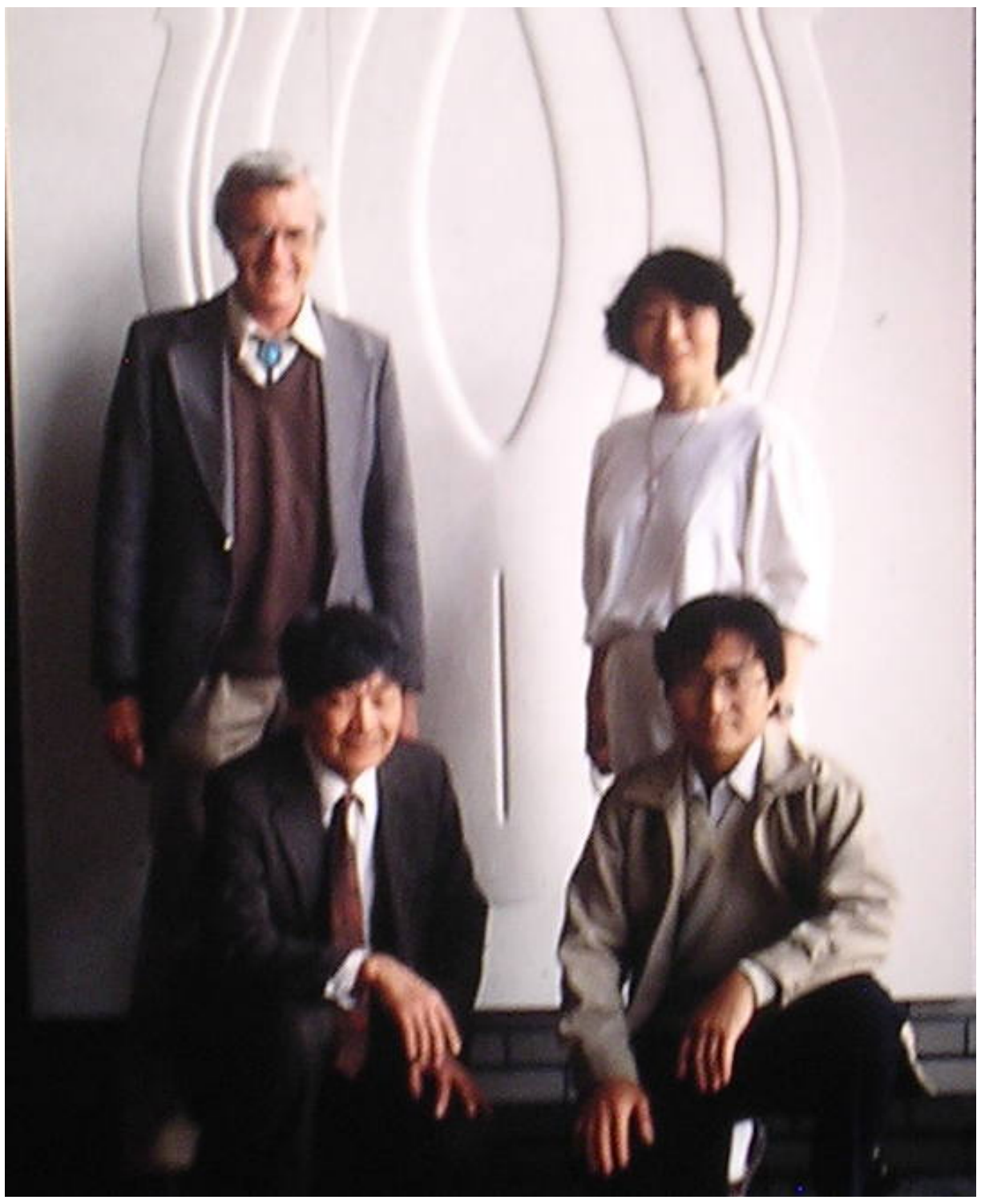

図3：新しい宇治研究所入口の彫刻の前。
Takekoshi, Iwashita, Jameson, Yokota-san
In front of the sculpture in
the new Uji Institute entrance.

（現広島国際学院大学）で数年教鞭を執られました。その後は自宅で菜園作りなどを楽しんでいらっしゃいました。先生のご冥福を心からお祈りいたします。

水本 元治（元原研）・岩下 芳久（京都大学）

**In Memory of Professor Hidekuni Takekoshi**

It is a pity that the former Atomic Energy Research Institute Senior Researcher, Professor Emeritus of Kyoto University Institute for Chemical Research and Kyoto University Hidekuni Takekoshi died on January 11 2020.

He was 93 years old. He was born and grew up in Taiwan.

Dr. Hidekuni Takekoshi was a student of Arakatsu Laboratory at Kyoto University at the time of the defeat, and participated in the reconstruction of cyclotron as a member of the former Arakatsu Laboratory after the war; the assembly was completed in 1955. In November of that year, the beam was extracted and the production of radioisotopes started. The building was the second commercial building of the 90-year-old Keage Power Station, which was the first commercial power station in Japan utilizing the water of Lake Biwa. It was an abandoned building and was leaking rain. It seems that he borrowed it from Kyoto City.

After that, from 1956, he joined the physics department of the former Japan Atomic Energy Research Institute (JAERI, now Japan Atomic Energy Agency). JAERI was the first in Japan to start research on the Mossbauer effect using thermal neutrons from nuclear reactors. From 1962, as the director of the electronic linac, he was instructed in the development, operation, and maintenance of accelerators and neutron nuclear physics using neutrons from bremsstrahlung from the electronic linac. The original JAERI electron linac had an acceleration energy of 20 MeV and was proud of the state-of-the-art performance in Japan at that time, but in response to the demand for higher performance, it was possible to increase the acceleration energy to 120 MeV from 1969 in about 3 years. (JAERI Report 1238, 1975). With this extremely limited budget, 17 staff members under the supervision of Dr. Takekoshi took charge of the production of the accelerator equipment, and almost all except the accelerator tube and the klystron were independently produced. Since then, much research and development was carried out, including the development of a high-power buncher using this accelerator, the development of a microwave window for a waveguide, and the development of a Ba-impregnated cathode for a large klystron. Also, in the production of experimental equipment, various measurement equipment was installed including the construction of 190 m neutron flight path etc. In producing the measuring instrument, Dr. Takekoshi himself produced the handmade device. The fact that the electronic linac was so-called "hand-made with Takekoshi color" was one of the factors that made it possible to improve the beam output and quality through continuous improvement and refurbishment. As a result, JAERI electron linac was able to use the beam for continuous operation and various researches for 21 years until 1993. Personally, Dr. Takekoshi is called “Uncle Takekoshi” (Mrs. Takekoshi was also in the same physical department), which is good for young researchers and means a high level of trust from technical support staff. He had a good reputation.

From 1976, he returned as professor to the Nuclear Science Research Facility (former Keage Branch) of the Institute for Chemical Research, Kyoto University, to support the cyclotron, which was devoted to research after the war. He carried out experiments on biological and material irradiation using cyclotron beams, research on superconducting cyclotrons, and the Kyoto University shared accelerator project. In order to improve the extraction efficiency of the cyclotron beam, modifications such as adding a magnetic channel to the extraction part were also made by him. He also realized a smear sample exchange carousel (detector is a GM tube bought from Aloka) made from a handmade special skill.

Iwashita was asked to configure the circuit and used the PC8801 etc. to control it.

Takekoshi also did arc welding etc. by himself. It is popular to wear only running shirts during the hot summer months. A protective mask for eyes did not cover his arms exposed to the UV rays from the arc points, which caused significant sunburn. This deepened our knowledge about arc welding.

As mentioned above, the building was from the former Keage Power Plant, so there was a remnant of a drainage channel underground, and the main channel was closed to form a pool with 80 tons of cooling water for the cyclotron. I remember that there was a small pond on the downstream side that was closed, and because of the measures against mosquito larvae, he released carps there and sometimes fed the crust of a slice of bread he bought at a bakery as food.

From around 1985, Professor Makoto Inoue worked on a plan to abolish the deteriorated cyclotron and move from the Keage district to the Uji district to construct an ion linear accelerator. On the day of his retirement, Professor Takekoshi was able to hand in the refund documents on March 31, 1990, which means that he was associated with the birth and termination of the Keage Cyclotron. ("Accelerator" Vol.3, No.4, pp.384-390, (2006), Vol.4, No.1, pp.18-23, (2007), Institute for Chemical Research "Obaku" No.29, 13-1)

When we evacuated the building at Keage, the carps were also collected; he moaned that his carps had been fished. His carps were released into a pond in the Uji area, but there is no such pond now. The building at Keage is currently managed by Kansai Electric Power Company, and it seems that it is open to the public from time to time.

Dr. R. A. Jameson of the Los Alamos National Laboratory, who had a close friendship, was staying at the time when the relocation plan began to move.

There were roughly 16 students trained at Keage, including 5 PhDs.

After retirement, a year later, he taught for several years at Hiroshima Denki University (now Hiroshima International Gakuin University). After that, he enjoyed gardening at home. We sincerely pray for the soul of the teacher.

Motoharu Mizumoto (former JAERI member), Yoshihisa Iwashita (Kyoto University)

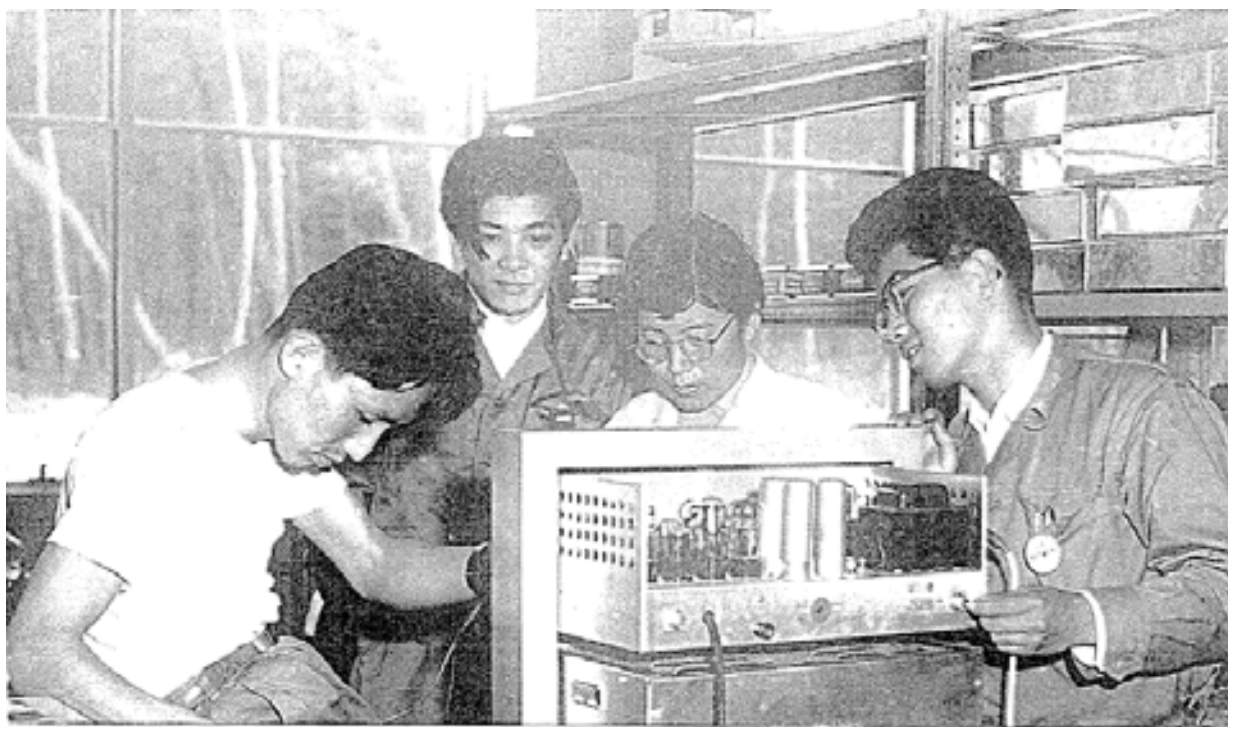

竹腰は左から2番目です。

Takekoshi is 2nd from left.

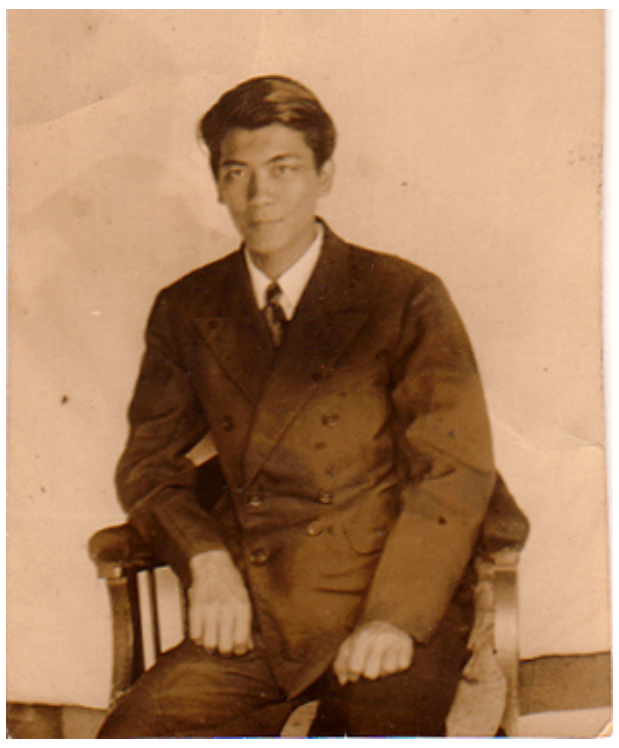

1949

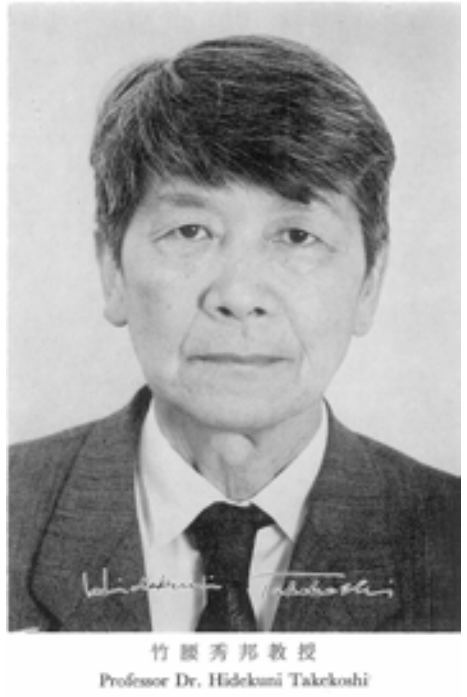

退職写真1990

Retirement photo 1990

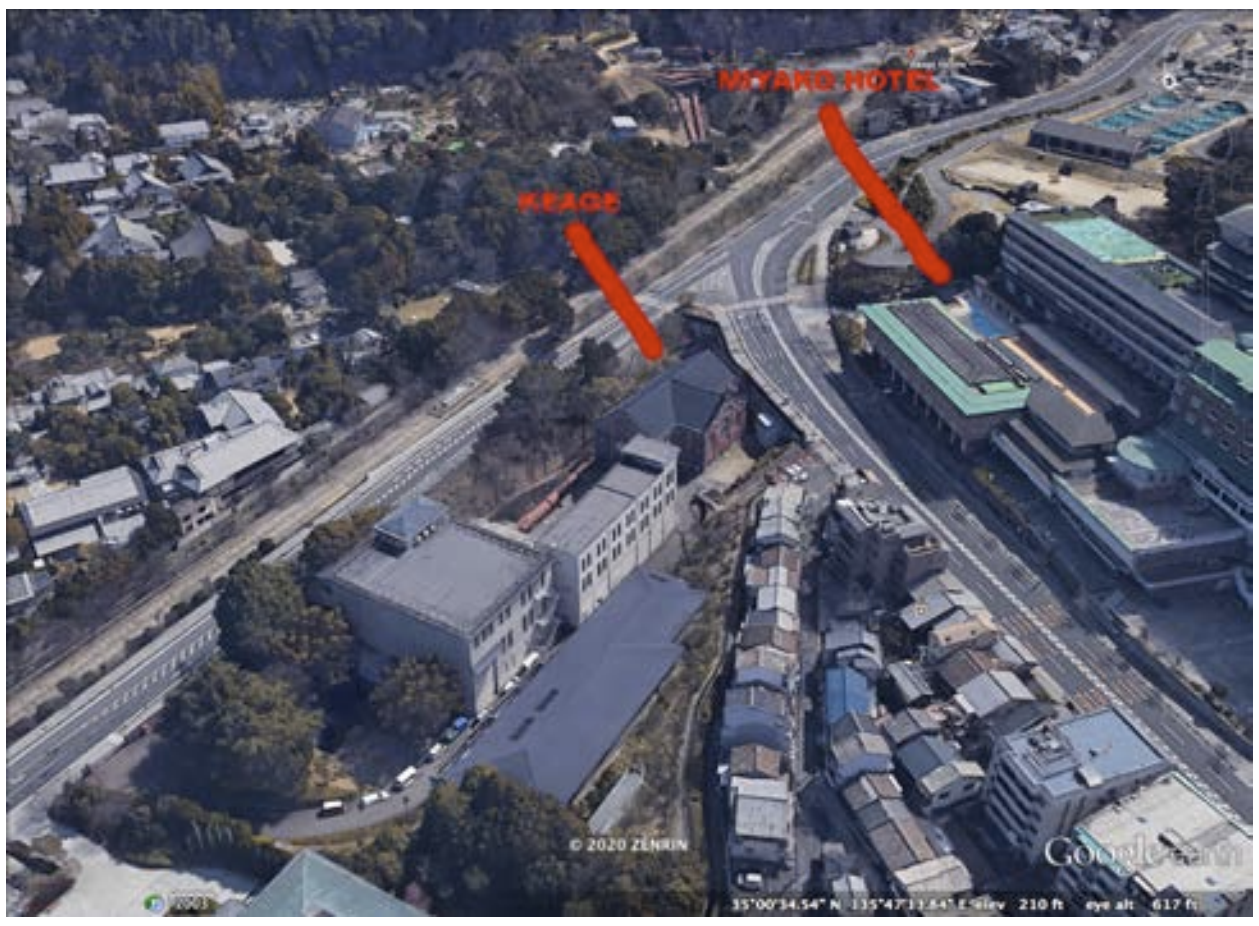

Keage

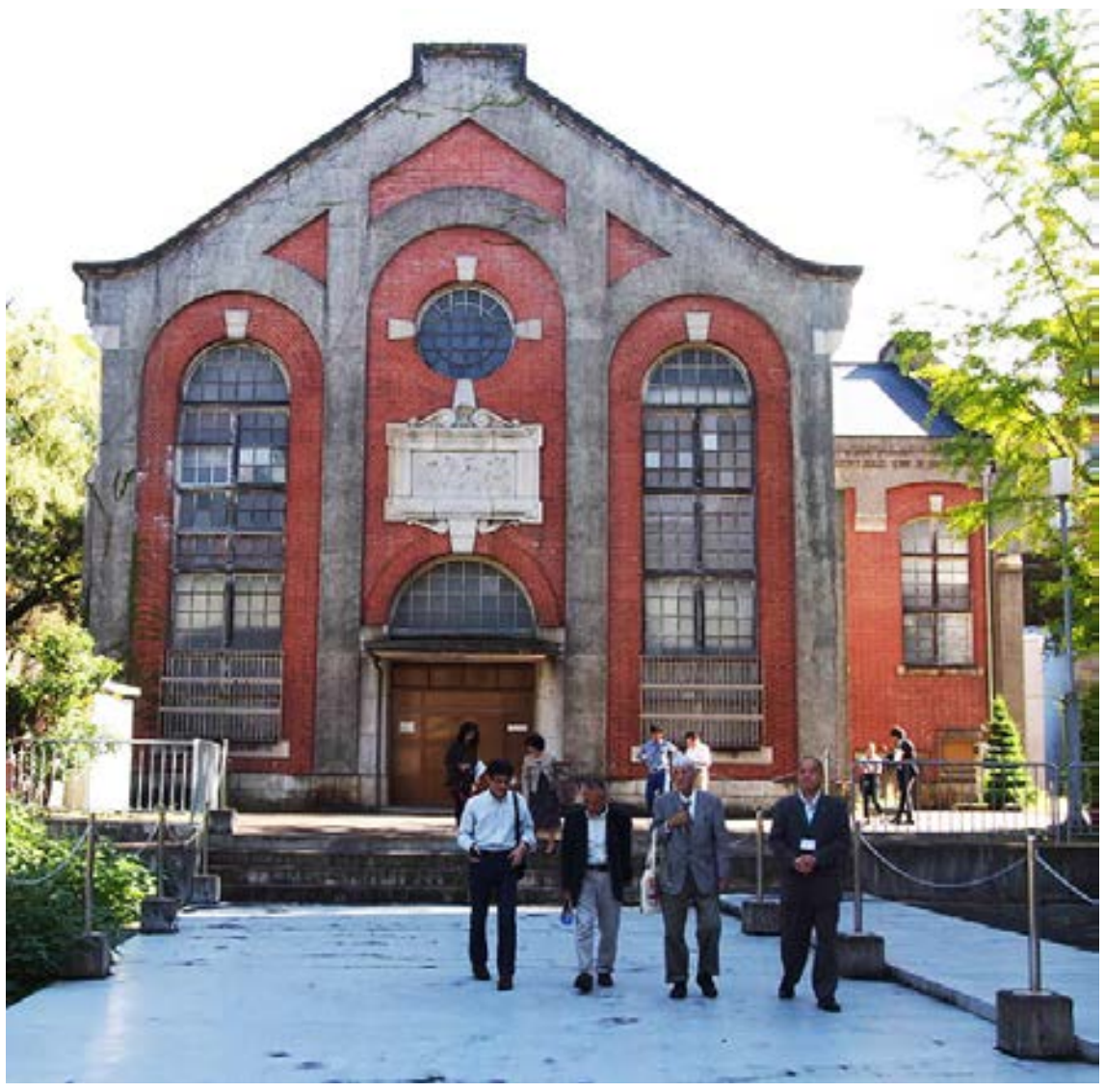

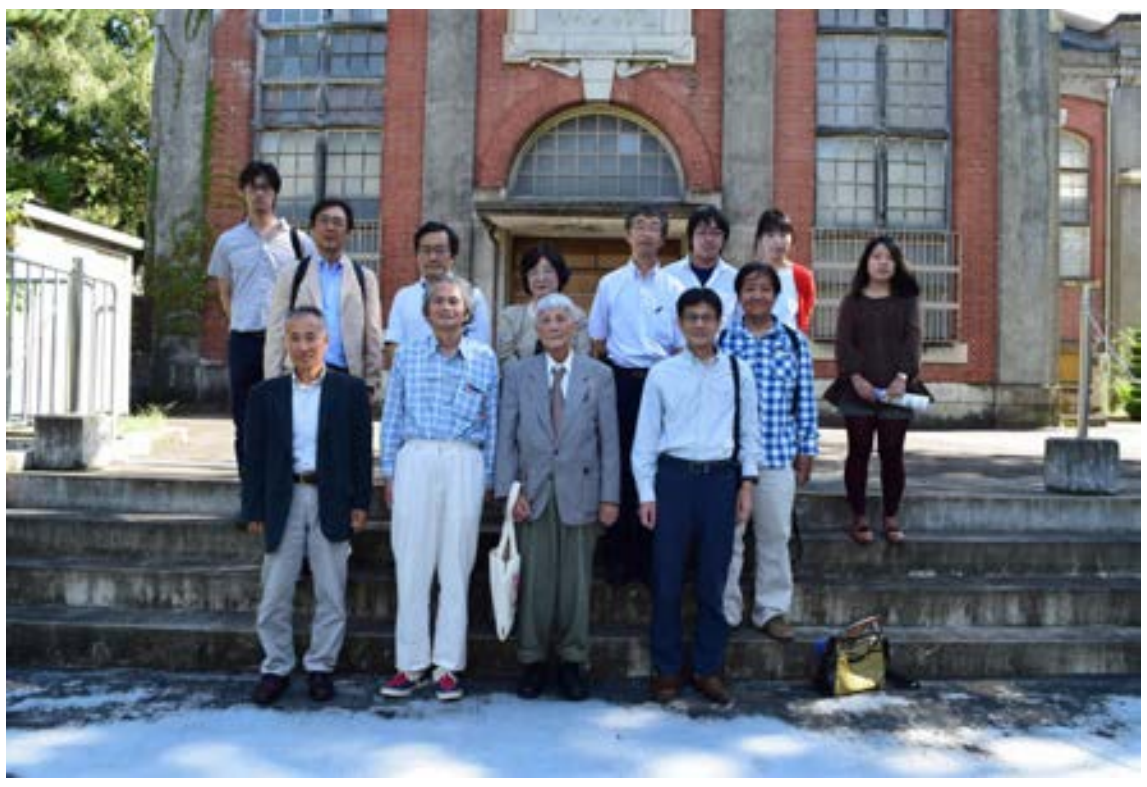

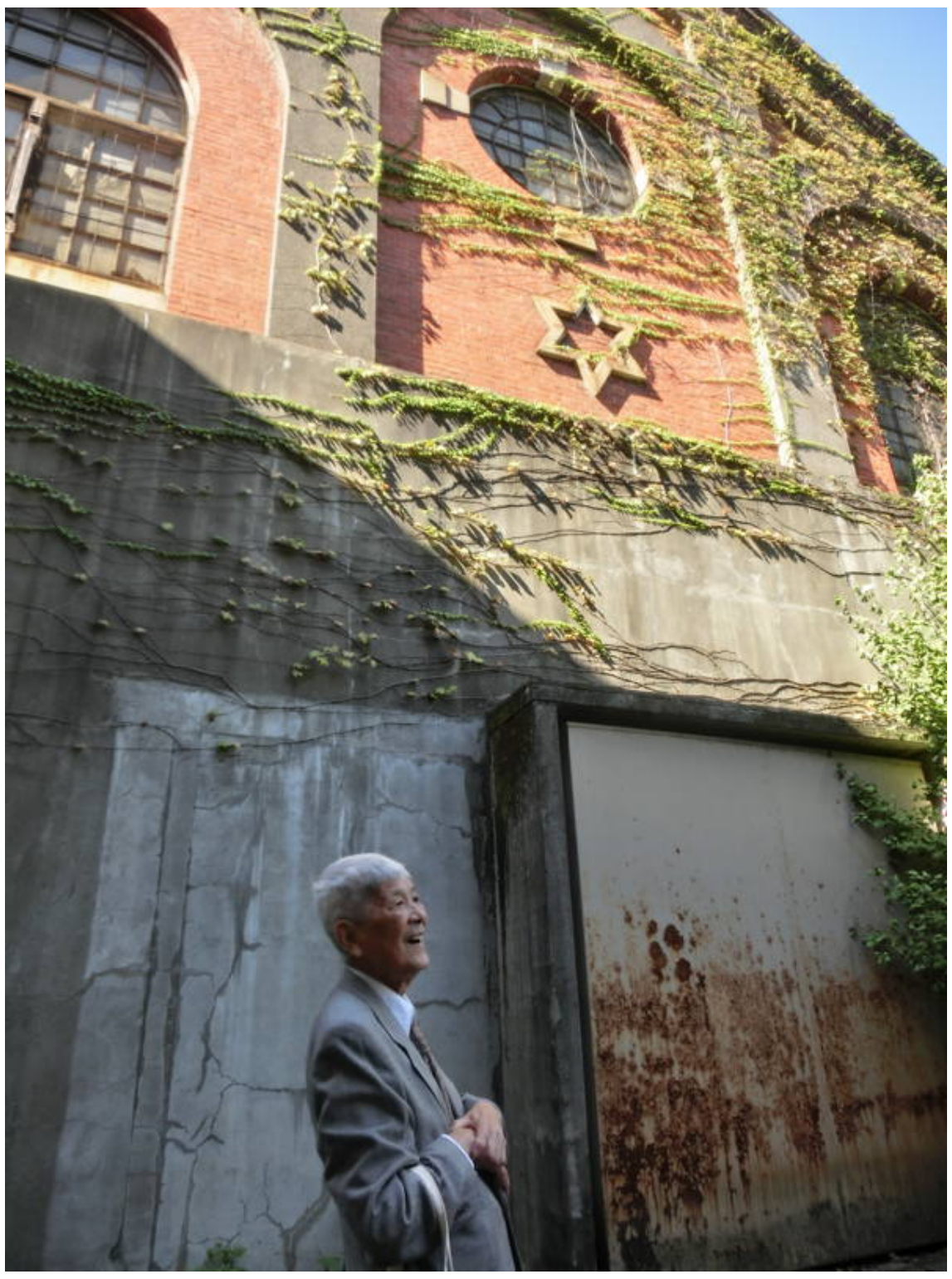

Keage Celebration 2015

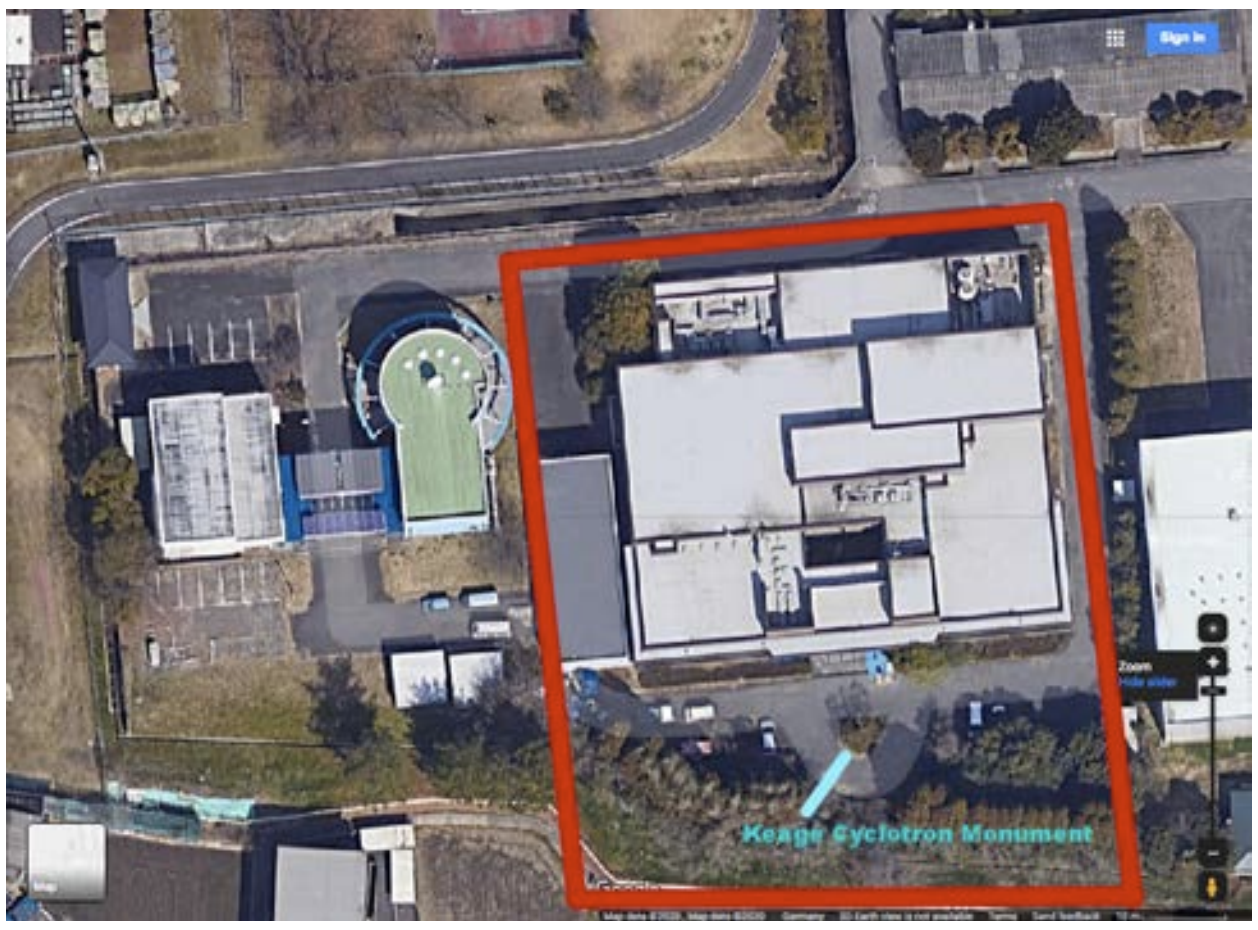

宇治キャンパスKU ICRアクセラレータラボ
Uji Campus KU ICR Accelerator Lab

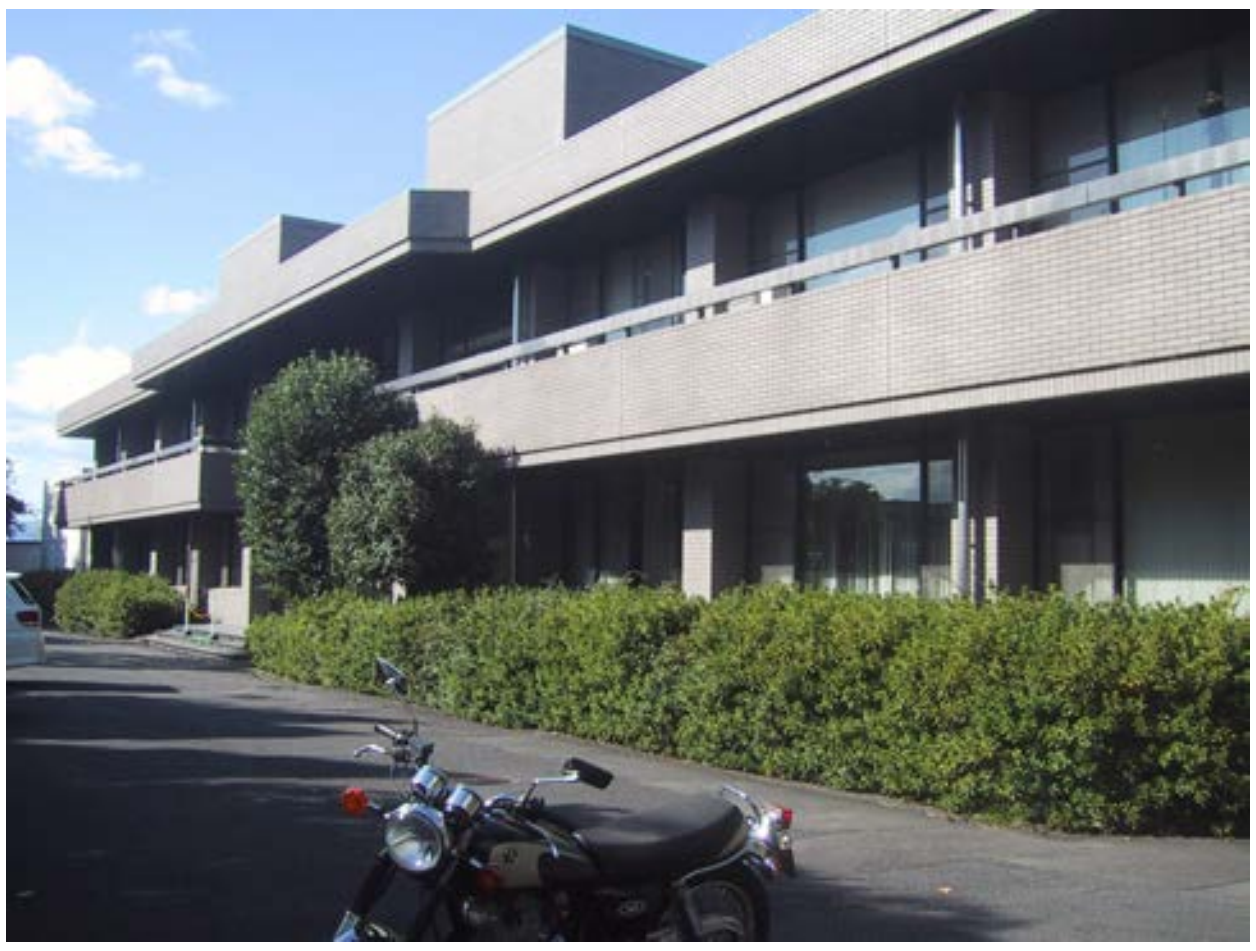

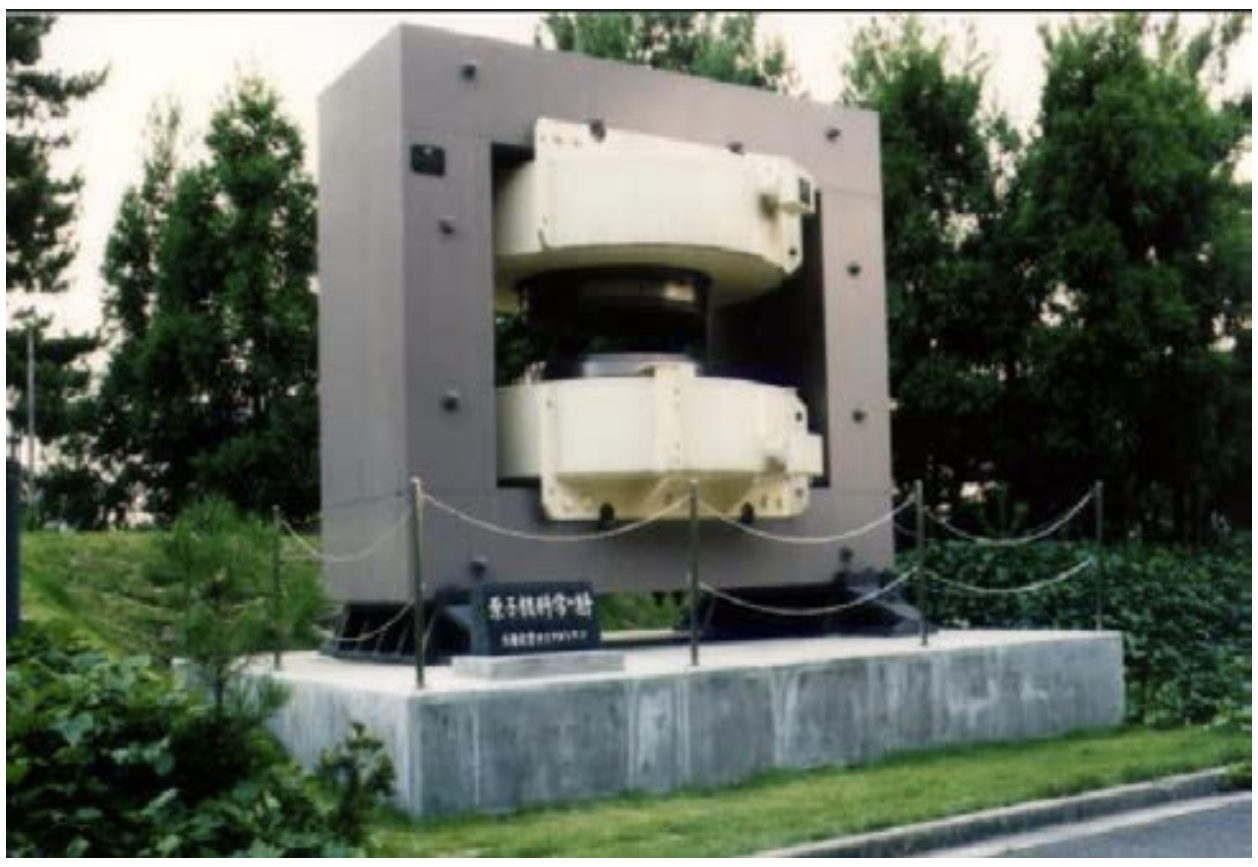

KU ICR宇治加速器研究所の蹴上サイクロトロン磁石記念碑
Keage cyclotron magnet monument at KU ICR Uji Accelerator Lab

# パーソナルメモリーズ– 水本元治
## Personal Memories – Motoharu Mizumoto

竹腰さんの思い出

私が最初に竹腰さんに会ったのは、1968年のことです。当時私は京都大学理学部原子核物理専攻の４年生でした。竹腰さんは日本原子力研究所（原研）のリニアックグループの室長で、大学へ学生のリクルートに来られたのでした。場所は理学部にあったタンデムバンデグラフの建屋の中でした。その頃の竹腰さんは40歳過ぎ、今となっては記憶が定かではありませんが、黒い硬い髪を少し額に長めに垂らしておられて、穏やかですが少しワイルドな印象のある方だったような気がします。私は家庭の事情もあり、大学院へは進学せず卒業したら就職することを決めており、すでに原子力関連の民間研究所から就職の誘いを受けていたのですが、竹腰さんからの話があり、一も二もなくそちらを断って原研を受験することを決心しました。就職試験の試験官の一人も、もちろん竹腰さんでした。幸い、竹腰さんからの口添えもあったと思いますが、入所試験に合格することができ、1969年の4月から原研のある茨城県の東海村で過ごすことになりました。

大学院に行かなかった私の研究者としての知識や技術は、研究所での経験や先輩方からの指導によって得られたものです。中でも竹腰さんからは最も大きな影響を受けました。学問的なこと、技術的なこと、さらには生き方についてのことなど、実に様々なことを教えて頂きました。その中で最も大きなものは、自分がやりたい実験の装置は自分で作るというものです。また、当時は、お金さえ出せば、出来合いの装置を購入出来るような状況でもありませんでした。しかし、リニアック研究室には、旋盤やボール盤などの工作用の機械、鉄、銅、アルミなどの様々な材料、真空管、トランジスターなどの電気部品などが充実していました。また、当時は計算機のマイクロプロセッサーなどが出始めたころで、世の中に先んじてそれらの使用法を試行錯誤で試している人たちもいました。さらに当時の高卒の技術サポートの人たちは日本各地から集められたかなり優秀で意欲溢れる人たちでした。そこに竹腰さんが室長ですから、様々な装置を自作するという点で鬼に金棒です。入所した次の年から始まったリニアックの増力計画ではこれらの資源を大いに活用してグループ一丸となってそのプロジェクトに当たったのです。

増力計画は極めて限られた予算で実行されました。また、その目標は20MeVだった電子のエネルギーを120MeVに向上させるという意欲的なものでした。必然的に加速管の長さが長くなりますから、加速器用の建屋を新たに増設しなければなりません。また、この装置の主目的は中性子飛行時間法による中性子断面積の精密測定ですから、長い飛行距離を有した測定室の建設も必要となります。結果的には190m、55m、40mと多くの施設が作られました。建物の建設費は節約が困難ですから、勢い加速器自体と測定装置は予算節約のため手作りが要求されます。その詳細は建設終了後、竹腰さんによってまとめられたJAERIレポート1238（1975）に詳しく記述されています。ここでは、その時に私が竹腰さんの指導の下に経験させて頂いた二つ例を述べたいと思います。一つは、安定化直流高圧電源の製作を命じられたことです。これは、クライストロン増幅器の前段のブースタークライストロン（SAS）の電源で、安定化は真空管で行うものでした。電圧は15kV位でそれほど大げさなものではなかったと思うのですが、大学卒業直後の新人にはそれ程容易な仕事ではありませんでした。背丈の半分くらいの高さの油つけトランスの上にアルミやベークライトの板を加工し、真空管回路を組み上げていくものです。ブリーダー抵抗は水道管として使用するものと同じ塩化ビニールのパイプの中に据え付けます。竹腰さんは当時エポキシ系の接着剤（アラルダイト）がお好みでよく利用されていましたが、ここでもそれを色々なところに用いて製作しました。そのようにして製作したものですから、今と違って安全に対する配慮に欠けるところもあり、電圧の印可試験では、竹腰さんからたびたび感電しないように十分注意するようにとの指示を受けたものです。この電源は、その後安全性を考慮したうえで、後々までリニアックの電源として利用されていました。

もう一方の検出器の製作では、私自身は大型の液体シンチレーション検出器の製作が主な仕事でしたが、竹腰さんが製作された全断面積測定用のリチウムグラス検出器

やウランのフィッションチェンバーの製作を手伝いました。傑作なのはリチウムグラス検出器の製作では11.1cmφのシンチレータを7個まとめて約40cmφ弱のアセンブリーを作るのですが、一つ一つの光電子増倍管のケースとしてお茶を入れる茶筒、前面の覆いには黒い絵の具を塗った厚紙を使用し、例によって、それらをアラルダイトで接着するというものでした。この装置は、原研の中性子測定の主力装置として190m飛行管の後に据え付けられ、長らく利用され様々な成果を上げました。

他の方々も述べられているように、当時独身だった私も竹腰さんの車で色々なところに連れて行ってもらいました。竹腰さんは身なりには全く無頓着で、原研の作業着のまま出かけられます。しかし、車にはこだわりがあったようで、かなり高級なスカイラインGTに乗っておられ、それも数年で新しいモデルに買い換えられていたように思います。車の運転はもちろん竹腰さんご自身がなさいますし、旅行に掛かった費用は全て竹腰さん持ちです。また、旅行には、なぜか、研究所の秘書の方など女性の方たちも一緒だったことが多かったと記憶しています。後年、竹腰さんに命じられてリニアックの同窓会を開催しこれら女性の方々も含めて当時一緒だった人達を招いて色々な思い出話に花を咲かせました。私は1975年に結婚したのですがもちろん仲人は竹腰さんご夫妻にお願いしました。結婚の翌年、私たち夫婦は米国のオークリッジ国立研究所の電子リニアック施設（ORELA）に約2年間滞在しました。多分1977年ころだと思いますが、オークリッジまで訪ねてこられた時、空港までお迎えに行きますとお伝えしたのですが、いつもの通りに人にはなるべく迷惑を掛けないとの信条から「必要ないよ、自分で行くから」とのこと、大型の空港リムジンを引き連れて通行ルートでも何でもない我々のアパートの前まで直接やってこられたのには驚きました。

1976年に京都大学に赴任されて以降もお付き合いは続きました。私は、その後1980年にリニアックでの中性子断面積の研究がまとまりましたので、竹腰さんにお願いして博士論文として大学に提出することになりました。何度か指導を受けるために、蹴上げの竹腰研究室を訪問することになります。その際、岩下芳久さんなどとのお付き合いが始まりましたし、竹腰さんに紹介してもらってBob Jamesonさんと初めてお会いしたのもその頃です。私が、加速器による放射性廃棄物の核変換計画（オメガ計画）の陽子加速器プロジェクトを任されたのは1988年ころからですが、Jamesonさんと彼の紹介で知り合った多くの陽子加速器の専門家との出会いが以降の陽子加速器開発に大いに役に立ちました。また、陽子加速器プロジェクトが始まってからは、竹腰さんの学生さん達に原研の実習プログラムへ参加してもらったり、竹腰研究室や竹腰研究室を引き継いだ研究室出身の優秀な卒業生の方々にプロジェクトに参加してもらいました。

以降、学会や研究会があるたびに竹腰さんのもとを何度も訪れました。京都はもとより、広島、東京などその都度相変わらず様々なところに連れて行ってもらったものです。後年は流石に車の運転を代わって私が行うこともありましたが、本当は最後までご自分で運転をやりたかったようで、何度か「運転うまくないね」などと言われたりしました。その後、直接竹腰さんが参加されなくなった後も、 Jamesonさんを始め竹腰さんを通して知り合った多くの方たちとの交流と様々な経験は私の人生の貴重な財産といえると思います。

一昨年以降体調を崩されてから二度ほど竹腰さんを病院に訪問しました。短い時間ではありましたが、原研リニアック時代の話をすることが出来ました。古い知人のことをいつまでも気にかけておられて近況を尋ねられたりしました。暖かくなったらまた、お伺いしようと思っていた矢先の突然の訃報でした。本当に残念でなりません。心からご冥福をお祈りします。

水本　元治

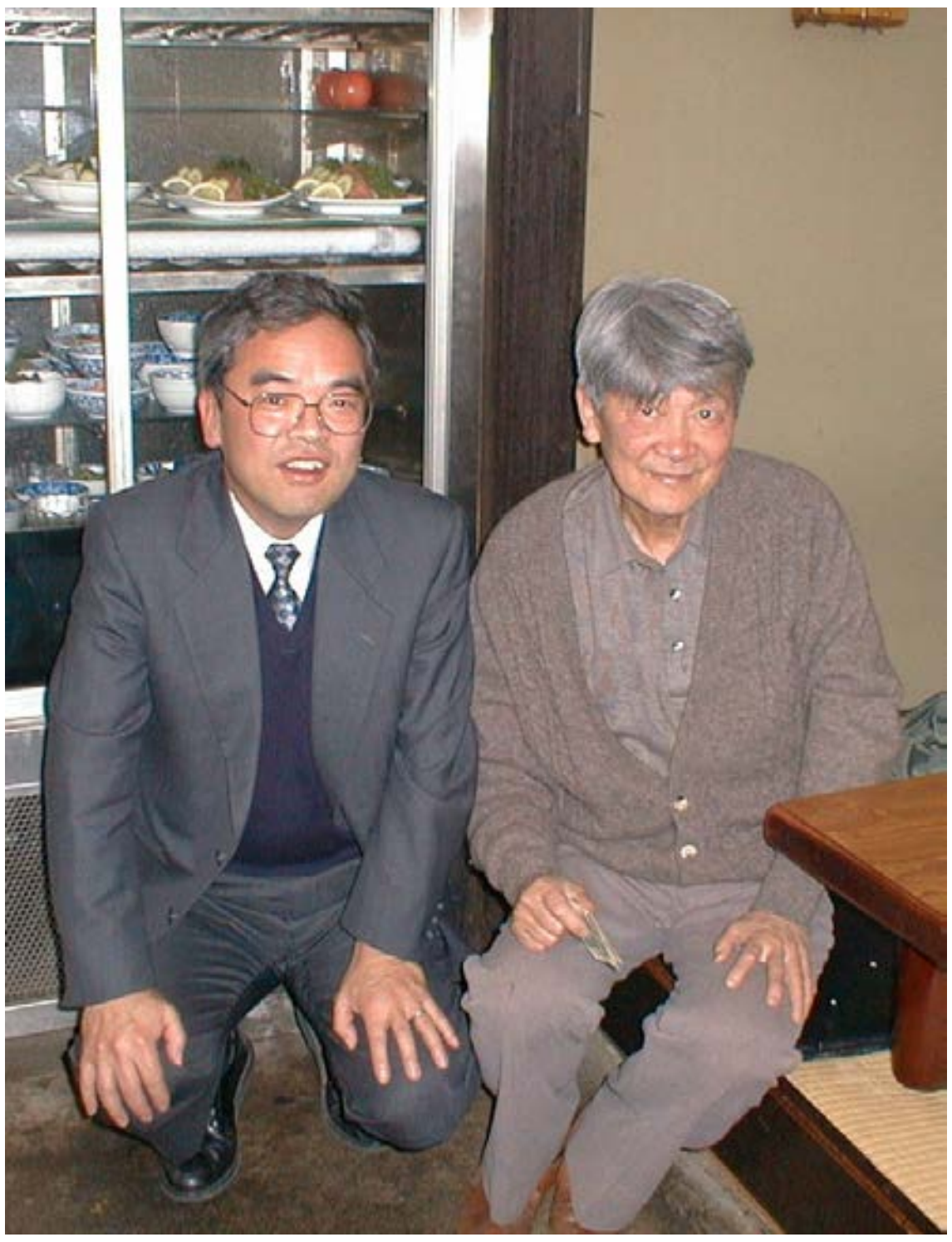

Mizumoto & Takekoshi – 1999

**Memory of Dr. Takekoshi**

Mizumoto Motoharu

I first met Dr. Takekoshi in 1968. At the time, I was a 4th year student at the Department of Nuclear Physics, Faculty of Science, Kyoto University. Dr. Takekoshi was the director of the Linac Group of the Japan Atomic Energy Research Institute (JAERI) and came to recruit students to the university. The place where I met him was in the building of Tandem Van de Graaff in the Faculty of Science. Dr. Takekoshi was over 40 years old, and it seems that he had a gentle but slightly wild impression with his black, hard hair hanging down a little on his forehead. Due to family circumstances, I decided to go to work after graduating without going on to graduate school, and I had already been invited to work by a nuclear-related private research institute. I decided to take the JAERI exam without a second thought. Of course, one of the examiners for the employment examination was Dr. Takekoshi. Fortunately, I think that Dr. Takekoshi had a chance to help me. I was able to pass the entrance examination, and from April 1969, I decided to spend my time at Tokai Village in Ibaraki Prefecture, where JAERI is located.

My knowledge and skills as a researcher, who did not go to graduate school, was acquired through experience at the institute and guidance from seniors. Above all, Dr. Takekoshi was the biggest influence. He taught me a variety of things, such as academic, technical, and even about life. The biggest of them is that you make your own experimental device. Also, at the time, it was not a situation where you could buy a ready-made device even though you could afford it. However, in the linac laboratory, there were plenty of machine tools such as lathes and drilling machines, various materials such as iron, copper and aluminum, vacuum tubes, and electrical parts such as vacuum tubes and transistors. In addition, at the time, when microprocessors for computers were coming out, some people were trying out their usage by trial and error ahead of the time. Furthermore, the technical support people of high school graduates at that time were quite excellent and motivated people from all over Japan. As the manager there, Dr. Takekoshi got an advantage to make various devices by his own way like a proverb saying "a powerful demon got an additional ion rod". In the Linac boosting plan that began the year after I entered the institute, the group worked together to make use of the most of these resources.

The boost plan was carried out with a very limited budget. The goal was also ambitious to increase the electron energy from 20 MeV to 120 MeV. Since the length of the accelerating tube would inevitably increase, it will be necessary to add a new building for the accelerator. Moreover, since the main purpose of this device was the precise measurement of the neutron cross section by the neutron time-of-flight method, it was also necessary to construct a measurement room with a long flight distance. As a result, 190m, 55m, 40m and many other facilities were built. Building construction costs are difficult to save; the accelerator itself and the measuring equipment must be handmade to save money. The details are described in JAERI Report 1238 (1975) compiled by Dr. Takekoshi after the construction. Here, I would like to mention two examples that I experienced under the guidance of Dr. Takekoshi. One is that he ordered me to manufacture a stabilized DC high voltage power supply. This was the power supply for the booster klystron (SAS) before the klystron amplifier, and the stabilization was done with a vacuum tube. The voltage was about 15kV, which was not too big, but it was not so easy for a newcomer who just graduated from university. Aluminum and bakelite plates were processed on an oiling transformer that was about half the height, and a vacuum tube circuit was assembled. The bleeder resistor was installed in the same PVC pipe used for the water pipe. Dr. Takekoshi often used epoxy adhesives (Araldite) at that time, so I also used it in various places. Since it was manufactured in this way, there were some points with lack safety consideration unlike present standard, and in the voltage application test, Dr. Takekoshi instructed me to be careful enough to avoid electric shock. This power supply was used as a power supply for the linac afterwards, after considering safety.

I was mainly involved in the production of large-scale liquid scintillation detector. At the same time, I helped Dr. Takekoshi for the other detector to make a lithium glass detector for measuring total cross-sections and a fission chamber of uranium. A masterpiece is the production of a lithium glass detector by combining seven 11.1 cmϕ (diam.) scintillators to make a round shape assembly of about 40 cmϕ (diam.). It also used a can for tea leaf container as case of each photomultiplier tube, and cardboard coated with black paint as a front cover and, as usual, glued them together with Araldite. This device was installed after the 190m flight tube as the main device for neutron measurement at JAERI linac, and was used for a long time with various results.

I was single at the time and Dr. Takekoshi took me to various places by his car. He was completely casual in appearance and could go out in the work clothes of JAERI. However, the car seems to have been particular to him, and a rather high-end Skyline GT, which was replaced with a new model in a few years. Dr. Takekoshi himself was responsible for driving the car, and responsible for all the travel expenses. I also remember that for some reason, many of the women who traveled with us, such as the secretary of the institute, were also on the trip. Later in the year, Dr. Takekoshi ordered me to hold a linac reunion and invite the people who were together at the time, including these ladies, to create various memories. I got married in 1975, and of course I asked Dr. and Mrs. Takekoshi to be the matchmaker. The year after we got married, our couple stayed for about two years at the Electron Linac Facility (ORELA) at Oak Ridge National Laboratory in the United States. It was probably around 1977. When he visited Oak Ridge, I told him that I will pick him up at the airport, but as usual he did not like to disturb people as much as possible. He said “No, I'm going to go by myself. " I was surprised that he managed to bring a large airport limousine directly to the front of our apartment, which was not a route.

The relationship between Dr.Takekoshi and me had been continued after he was transferred to Kyoto University in 1976. In 1980, my research work on neutron cross section at the Linac was completed, so I asked Dr. Takekoshi to submit it to the university as a doctoral dissertation. I visited the Keage Takekoshi Lab at Institute for Chemical Research (ICR) several times for his guidance. At that time, I started to interact with Dr. Yoshihisa Iwashita, and that was when I first met Dr. Bob Jameson after being introduced by Dr. Takekoshi. It was around 1988 that I was assigned to the proton accelerator project of the accelerator-based nuclear waste transmutation project (Omega project). The encounter with Jameson and many proton accelerator experts with his introduction was extremely useful for the subsequent development of the proton accelerator. Also, since the proton accelerator project started, Dr. Takekoshi's students participated in the practical training program of JAERI, and excellent graduates from the accelerator laboratory of ICR joined the project.

Since then, I visited Dr. Takekoshi many times when there were academic conferences and research meetings. In addition to Kyoto, he took me to various interesting sites such as Hiroshima, Nara and Tokyo. In later years, I sometimes took over driving the car as a matter of course, but in reality it seemed that he wanted to do it himself until the end, and I was repeatedly told that “you are not a good driver”.

Even after Dr. Takekoshi could not participate directly with us, the exchanges and various experiences with Jameson and many other people I met through Dr. Takekoshi are valuable assets of my life.

After he became ill since last year, I visited Dr. Takekoshi twice at the hospital. Although it was a short time, we were able to talk about the days of the JAERI Linac. He was always worried about his old acquaintances and asked about their current situation. It was a sudden obituary although I had been thinking of visiting again when it became warm. I am really sorry and would like to express my deepest sympathy to Mr Takekoshi.

Motoharu Mizumoto

## Personal Memories – Takao Inagaki

竹腰先生、安らかに

春を迎え気持ちの良い青空のなか、自宅近くの山をのんびりと歩いているとザックの中の携帯が鳴りました。宇治の研究室からです。普段ならメールでの連絡のはずですから、一瞬、嫌な予感がしました。竹腰先生の訃報でした。しかし、信じられません。家に帰りパソコンでメールを確認すると、亡くなったことを伝える文書が届いていました。実は、亡くなられる二日前に、入院中の先生にお会いしたばかりでした。また会いに来ますよと言って帰ったのに、それが、最後の別れになってしまいました。

時計は四十五年近く前に戻ります。私が竹腰先生のお世話になったのは、1977年

からの原子核工学科の修士課程の二年間でした。先生は1976年に化研蹴上に赴任されましたので、私は二代目の学生です。京大に戻られてすぐの頃で、これからの研究をどう進めて行こうかと探っておられる時だったのではと思います。私は、加速器を使った応用ということでPIXEの実験を始めることになります。実はその前にチャネリングの実験に挑戦しようとしたのですが、自作のゴニオメーター（とてもゴニオと呼べない代物でした）では、全く位置精度が出ず、チャネリングのチャの字も起きません。先生に相談して、別のテーマPIXEに方向転換することになりました。PIXEの実験では、先生はいろいろと変わった測定試料を持って来られました。一番の変わり種は、マウスのガン組織です。いわば肉片です。測定するためにはこの肉片を薄くスライスしなければなりません。当時の私はガンと聞いただけで、手が出ず、結局、先生に下準備をしていただく始末でした。その頃の先生は、このような加速器の工学的、医学的な応用、一方では、現在の加速器の改良、新たな加速器の構想に向けて走っておられたと思います。コイルを作るため、学生総出で、屋外で太い銅線にガラスウールで被覆し、その銅線を旋盤でコイルに巻いたりしたことを覚えています（このガラスウールは皮膚に刺さり、後でチクチクして大変な思いをすることになります）。ちょうどこの頃から、先生と超電導コイルの勉強も始めました。私は、修士課程の二年間だけお世話になり、その後は、卒業して民間企業に就職することになりました。そこで最後の課題が竹腰先生から出ました。修士論文を英語にして化学研究所の年報に載せなさいというものです。今後、研究者として論文を書くことはないだろうから、論文に仕上げなさいというものです。しかし、これは私にとっては難問で、英語との戦いです。寒い蹴上の研究室を抜け出し、向かい

の都ホテルのロビーに竹腰先生と何時間も居座って論文作成と格闘しました。が、三月末になっても完成せず、未完の論文を残したまま、私は就職のため東京に行ってしまいました。何ヶ月かして先生からお呼びがかかります。残した論文を何とかしろというものです。出張にかこつけて、蹴上研究室に戻り、さらに休暇も取って、論文の体裁を整えたように思います。竹腰先生には、ずいぶん辛抱して、何とか掲載されるまで、私の論文の面倒を見ていただいたとたいへん感謝しております。

時計はずっと戻って十五年ほど前です。竹腰先生は、八十才近くになっておられます。蹴上の研究所では、地下、天井裏に何か使えそうなものはないかと探し回って、いろいろな実験装置を手作されていましたが、八十才近くになってもその意欲は衰えず、ある時、先生から電話がありました。パソコンにボードを挿してセンサーからデータを取り込みたい、プログラムはどうしたらよいのかというものでした。ホームセンターで部品を買い買い込み、パソコンにつなげて実験装置を作っておられました。いろいろなことに興味を持ち、さらにそれを実行される先生のバイタリティにはただただ敬服するばかりです。

時計は現在に戻ります。数年前に竹腰先生の米寿、卒寿のお祝いをした時には、先生は足腰もしっかりとしておられ、大変お元気な様子で、次の白寿も元気に迎えていただけるのではと思っていました。しかし、その後、体調を崩されてから、病院、介護施設での生活を余儀なくされました。病院では、看護師の皆さんにも「先生、先生」と呼ばれ、慕われておられました。最後に病院でお会いした時も、私が誰かはっきりと分かっておられ、また来ますよとお別れした時もしっかりっと返事を返されていたのに、まさかその二日後に亡くなられるとは、信じられない思いです。竹腰先生は、私の父と同じ大正十五年生まれで、私の父親のような存在でした。太平洋戦争が終わった時には十九歳の学生で、これから核物理研究の道を歩まれるスタートラインに立たれた時だったと思います。それからの先生は、戦後の日本の加速器とともに歩まれた、まさに日本の加速器の歴史の生き証人だったのではないかと思います。竹腰先生のご冥福を心からお祈り申し上げます。ありがとうございました。

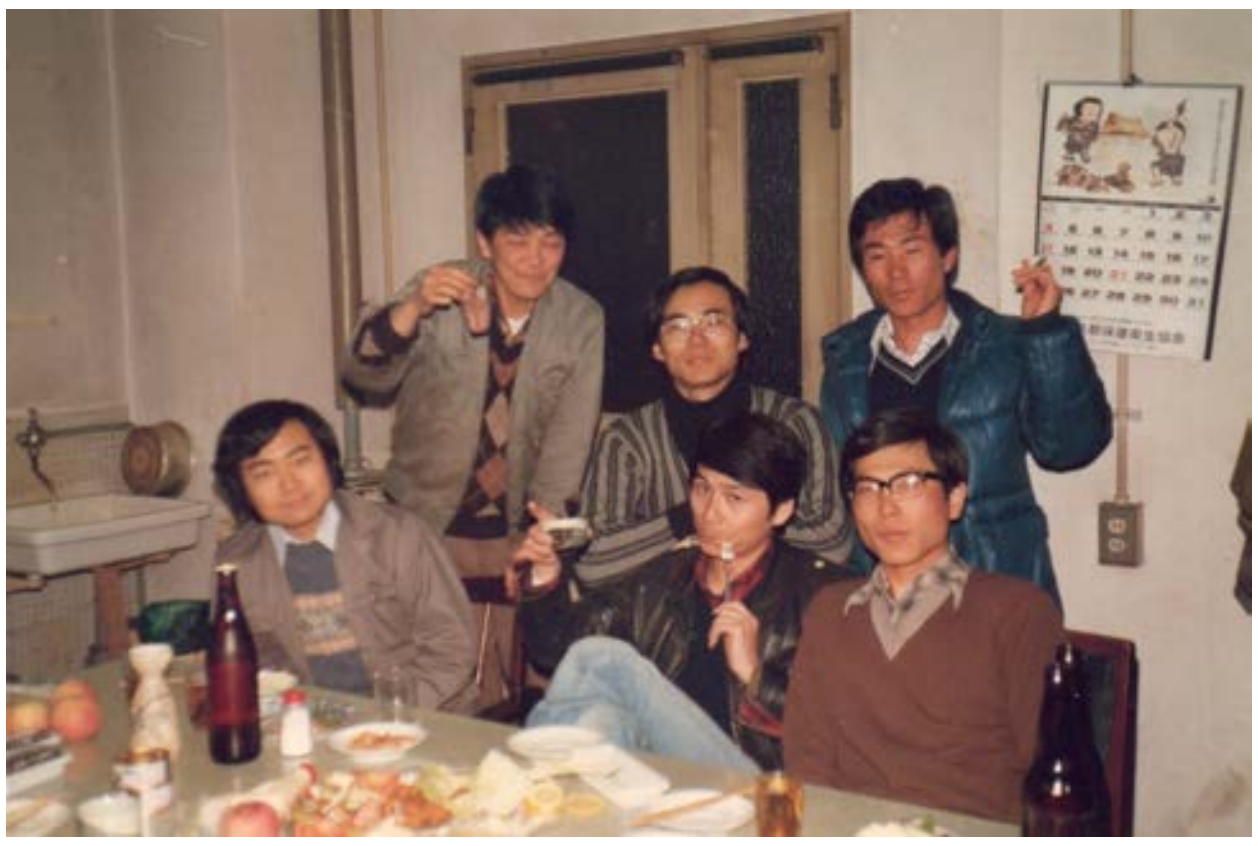

1979年の送別会–右に立っているいがき、左に立っている竹腰、左に座っている岩下

Farewell party 1979 – Igaki standing right, Takekoshi standing left, Iwashita sitting left

**Takekoshi sensei, rest in peace**

Takao Inagaki

The mobile phone in my backpack rang when I was leisurely walking in the mountains near my home under the pleasant blue sky in spring. The call was from the laboratory on the Uji campus of Kyoto University. I got a bad feeling in that moment because e-mails are the regular contact means between me and the lab. It was the obituary of Professor Takekoshi, but I just couldn't believe it. I went back home, checked e-mails on my computer, and found the message informing me of his death. He passed away just two days after I met him at the hospital. When leaving, I said "I'll be back to see you again", which became the last words I spoke to him.

The clock turns back almost 45 years ago. I was taken care of by Professor Takekoshi during the two years of the Master's program in the Department of Nuclear Engineering from 1977. He became a professor at the Keage lab of ICR (Institute for Chemical Research, Kyoto University) in 1976, which means that I am a student of his second year in Kyoto. As he had just started to work at a new place, I think that he was still in the middle of looking around for the research subjects he should study. There was an accelerator (cyclotron) at Keage, so I eventually did an experiment study using the PIXE (Particle Induced X-ray Emission) technique. To tell the truth, I first tried channeling experiments using a goniometer I made myself, but my machine was far from a real goniometer.... very low precision of position measurement. It was even impossible to observe channeling. After consulting with Prof. Takekoshi, I decided to change direction to another theme, PIXE. In the PIXE experiment, he brought me various unusual measurement samples. The most unique sample given was mouse cancer tissue, a piece of meat, so to speak. It must be thinly sliced before measurement. Shamefully, I was afraid of even just touching “cancer”, so my supervisor had to prepare the sample for me. I think that, at that time, Prof. Takekoshi’s main research interest was directed to the engineering and medical application of particle accelerators. He was, on the other hand, interested in the improvement of the cyclotron and even elaborating a plan to construct a new machine in Kyoto.

I remember that all the students worked outdoors to make a coil. We diligently covered a thick copper wire with glass wool and and wound it into a coil on a lathe. (We later had a hard time because our skin kept prickling against the glass wool remained on the clothes.) Around this time, I started studying superconducting coils with Prof. Takekoshi. I only took the master's course for two years, after which I decided to graduate and work for a private company. Then I got the final homework from him, that is, “translate your master’s thesis into English and publish it in the annual report of ICR. He assumed that it would be the last chance for me to write up a physics research paper in English. This was, however, quite a tough job – I had to fight against English writing. Since it was pretty cold inside the Keage lab, we walked out and crossed the Sanjyo street to the Miyako Hotel. In the lobby, sitting beside Prof. Takekoshi, I struggled for hours to complete the English version. The homework was, however, not finished by the end of March. I moved to Tokyo to work for the company, leaving an unfinished paper behind. A few months later, he called me and said “do something about your paper”. On a business trip, I returned to the Keage lab, and even took a vacation to get the paper done. I am very grateful to Prof. Takekoshi for his patience and a great deal of effort that led to the publication of my paper in the end.

The clock returns to the present. When I celebrated Prof. Takekoshi's 88th and 90th birthday several years ago, he had much energy and no problem walking around. He looked very well, so I thought that he would reach his 99th birthday. However, shortly after, he became ill and was forced to live in a hospital and nursing home. He was liked by the nurses there who called him “sensei” which means “teacher” in Japanese.

The last time I met him at the hospital, there was an article about the accelerator next to his bed; he told me about it, and gave me a clear reply when I said goodbye. But only two days later, he had passed away, which is difficult for me to accept. Prof. Takekoshi was born in 1926 (Taisho era), the same year as my father was born. He was like my father. He was a nineteen-year-old student at the end of the Pacific War and just about to make a fruitful career as a physics researcher. I think that since then, he was exactly a living witness of the history of Japanese accelerators, who had actually moved forward together with post-war accelerators in this country. I pray sincerely for the repose of his soul. Prof. Takekoshi, thank you very much.

**Takao Ogaki**

# Personal Memories – Hiromi Okamoto

竹腰先生との思い出

私が竹腰グループに加わったのは1984年の春でした。当時はまだ蹴上に研究室があり、目の前の三条通りを路面電車が走っていました。通りを挟んで向こう側に豪華な都ホテルがそびえ立っていて、我々のいる古い建物を見下ろしていました。京都大学の本部から離れていることもあって学生の数は少なく、私は先生の指導の下で博士号を取得した最後の大学院生です。先生は私が学位を得た翌年に定年退官されました。
私にとって竹腰先生は“指導教員”というより“親父さん”のような存在でした。懐かしく思い出されるのは研究上のあれこれよりもむしろ、一緒に飲んだり、遠出したり、喧嘩をしたりした時のことがほとんどです。仕事に関連した話題は別の方が取り上げるでしょうから、私は先生の姿を思い浮かべながら、たわいもない個人的な昔話に終始したいと思います。

京都大学の物理学第二教室には通称“タコ部屋”というのがあって（今も存在するのかどうかは知りません）、修士課程１年時は全ての研究室の院生がそこで過ごすことになっていました。竹腰研は遠く離れた蹴上が本拠である上、私は同期の精鋭達とは違って日々遊び回っていたので、最初の１年間はほとんど先生と顔を合わせることがありませんでした。修士１年が終わりに近づいたある日、下宿に電報が届きます。電報なんて受け取るのは生まれて初めてで、一体何事かと驚いたのですが、文面は「早急に相談したいことがあるので研究室に出て来てもらえませんか　竹腰」でした。私は貧乏学生で下宿に電話がなく、蹴上にも滅多に行かなかったので、電報で呼び出す以外に方法がなかったわけです。並の教授ならば、こんな不肖の弟子とは早々に縁切りを考えるところでしょうけれども、私にとって幸いなことに竹腰先生は違いました。好き勝手をやっている私を温かく見守って下さいました。ちなみに、先生から電報を受け取ったことがもう一度だけあります。何年生の時だったかは忘れてしまいましたが、柄にもなく体調を壊してしまい、１週間以上も無断で研究室を休んだことがありました。この時もらった電報には「ここのところ研究室で見かけませんが、どうしていますか　竹腰」と書かれており、指導教員に心配をかけていたのでした…。
修士２年生になって以降は蹴上の研究室に毎日（？）顔を出すようになりました。そこにあった大昔のサイクロトロンに代わって新型の低エネルギー陽子線形加速器をつくる計画が認められ、私も開発メンバーに加わりました。稼働周波数を433.3 MHzと非常に高いところに設定し、当時最先端の技術を積極的に導入することになりました。私が最初に与えられた研究テーマは永久磁石四重極（PMQ）レンズの設計に関するもので、レンズに組み上げる前の磁石片の磁化方向測定装置を先生と一緒に製作しました。多重極磁場の測定システムも自作したのですが、こちらは柳父研の助手だった岩下芳久さんに手伝ってもらいました。大学院生は私以外に沢村勝さんがおり、彼は主にアルバレ型DTLのポストカップラーの研究をやっていたと記憶しています。このとき組み上げたPMQは現在京都大学宇治キャンパスにあるDTLのドリフトチューブに収まっています。
比較的早く修論を書き上げて時間をもてあまし気味だった私は個人的に興味のあったRFQリニアックに関する文献を片っ端から読んでいました。いまやRFQは定番のイオン入射器となっていますが、当時の日本に実用機はまだ存在しませんでした。周りに専門家がいないので、主に米国の加速器屋が書いた論文を通じて基本的なところを勉強していたわけです。ある日、先生に呼び止められ「今度つくるRFQの基本設計なんだけど岡本さんやってよ、君が一番詳しそうだから。」といきなり言われてしまいました。駆け出しの院生にこんな大事を頼むなんて無茶な話ですが、竹腰先生らしいと言えばらしい決断です。もう一人、別の先生もPARMTEQコードでシミュレーション計算をしていたはずですが、結局、私の提案したパラメータが採用されました。
ミーティングでRFQのベーン電極のモデュレーション加工法か何かについて発表した時のことです。先生が「その辺のところは適当で構わん」という感じの発言をし、カチンときた私が「適当でよいわけがない」と言い返して口論になりました。最終的に私は「そんないい加減なことを言うのなら、後の計算は自分でやって下さい」と捨て台詞を残して、プレゼンを終わりました。ミーティングの後、少し間を置いてから先生が私のところへやって来て、柔らかな表情でこう言いました：「さっきは怒ってたね...今晩、どこかで一杯やろうか」
蹴上研究室の真ん前には路面電車の停留所があり、その気になればすぐ河原町三条に繰り出すことができます。先生には頻繁に飲みに連れて行ってもらいました。岩下さんや沢村さんも一緒のことが多く、ま

ず居酒屋で楽しんだ後、京阪三条に近いお洒落なスナックに足を運ぶこともよくありました。その間、私のような貧乏学生はびた一文払わなくてもよいのです！先生と飲みに行ってお金を払った記憶はありません。

加速器技術関係の研究会へは、旅費節約のため、先生の車で現地へ向かったものです。学生達を乗せ、夜中に高速道路を走るのですが、ほとんど先生が運転していたように記憶しています。少なくとも私は一切運転していません、博士課程の途中まで免許を持っていなかったので。眠くなるとよくチョコレートを頬張ってらっしゃいました。さらに疲れてくるとハンドルの上で手を組み、その上にあごを乗せて運転し始めるのですが、カーブに合わせてハンドルを切っているのか、無意識に船をこぐタイミングとカーブがたまたま一致して難を逃れたのか分からず、助手席で緊張していたのを覚えています。

博士後期課程でも引き続き陽子線形加速器のR&Dに携わったのですが、博士論文はそれとは全く無関係のテーマで書きました。DTLの勉強をしているうち交番位相収束（APF）に興味がわき、その理論的側面を掘り下げることにしたのです。誰かに勧められたわけではなく勝手に始めたのですが、非常にマイナーなタイプの加速構造で、手に入る文献のほとんどはロシア語で書かれていました。ロシア語はちんぷんかんぷんで困っていると、竹腰先生がどこからか辞書を引っ張り出してきて、私のためにいくつかの文献を翻訳して下さいました。先生が休憩室で、辞書を片手にロシア語と格闘している姿が目に浮かびます。

私が博士論文の仕事を進めていた頃、先生と親交のあったBob Jamesonがサバティカルで来日し、暫くの間京都に滞在しました。竹腰先生とBobの親交は先生が亡くなるまで途絶えることなく続きます。BobからはAPF関連のロシア語論文の英語訳をいくつか頂いた他、私の書いた博士論文に有益なコメントを頂いたりもしました。先生を通じてBobと知り合ったことは、その後の私の研究屋としての人生に大きな影響を及ぼしました（もちろん、プラスの影響です）。

博士号を取得した私はBobの勧めでアメリカへ渡りました。長期間アメリカで過ごす覚悟だったのですが、蹴上で助手ポストが空き、結局そこへ収まるため半年ほどで帰国することになってしまいました。私を乗せた帰国便は夜遅く大阪の空港に到着したのですが、竹腰先生が出迎えに来てくれていました。車で京都に向かう途中、「日本食が恋しいだろう」と言われ、深夜のレストランで食事をご馳走になりました。

その2年後、私は京都で結婚式を挙げました。言うまでもなく、仲人は竹腰ご夫妻です。披露宴にはBobの姿もありました。媒酌人挨拶では「学生の頃、岡本君は毎日午後2時頃研究室にやって来て、夕方6時にはもう消えていた」など、ぼろくそに言われたのを憶えています。その時は別段何とも思いませんでしたが、この際一言だけ言わせて頂きます：「先生、“午後2時”は言い過ぎです。正午頃には出勤していましたよ！」

定年後しばらくして、先生は広島の私立大学の教授に就任されました。お誘いを受けて、広島のマンションに一度お邪魔したことがあります。先生は土いじりがお好きで、部屋の中に観葉植物や植木鉢からあふれた泥などが散らかっていました。研究室にも案内されたのですが、工学部の学生相手にスターリングエンジンを自作して楽しんでらっしゃいました。将来のエネルギー問題を見据えて太陽電池にも注目されており、既に研究用のサンプルを入手済みだったと思います。その日の晩は市内の繁華街で飲み屋をはしごし、遅くにマンションへ戻って、植物の間で寝ました。

1998年の秋、私は住み慣れた京都の街を離れ、広島に引っ越しました。引っ越した時点で先生はまだ例の私立大に籍を置いていました。確か、翌年の春に教授職を退いて持ち家のある京田辺へ戻られたはずなので、半年近くの間、車で１時間かからない所に住んでらっしゃったことになります。その間、取れたての海の幸を私のいた官舎まで届けてくださったり、一緒に遠出することも何度かありました。まだ１歳だった私の長男を喜ばせるため、動物とのふれあいが楽しめるテーマパークに連れて行ってくれた時のことは私よりも妻の方がよく憶えているようです。ベビーカーを押していた妻を置き去りにして私がすたすたと先へ歩き去った後、先生が全ての荷物を抱えて付き添ってくれたそうです。ところで、この頃、大学ではソーラーカーのモデルが完成していたはずです。

京田辺に戻られて以降は、先生と直接お会いする機会がほとんど失われてしまいました。ですが、時々思い出したかのように差し入れが郵送されてきました。私の方からもお盆や年末辺りに電話を入れてご機嫌伺いをしていました。その電話も数年前から突然つながらなくなりました。いつのことだったか思い出せないのですが、出張か何かで京都へ出向いた際、二人で駅前の居酒屋に入りました。私が電話をかけて先生を呼び出し、例によって飲み代をたかったわけです。先生とお会いしたのはその時が最後です。何年前だったか全く思い出せないのに、座った席や店の雰囲気、注文した料理の一部、そして何より先生の笑顔は

はっきり記憶に残っています。

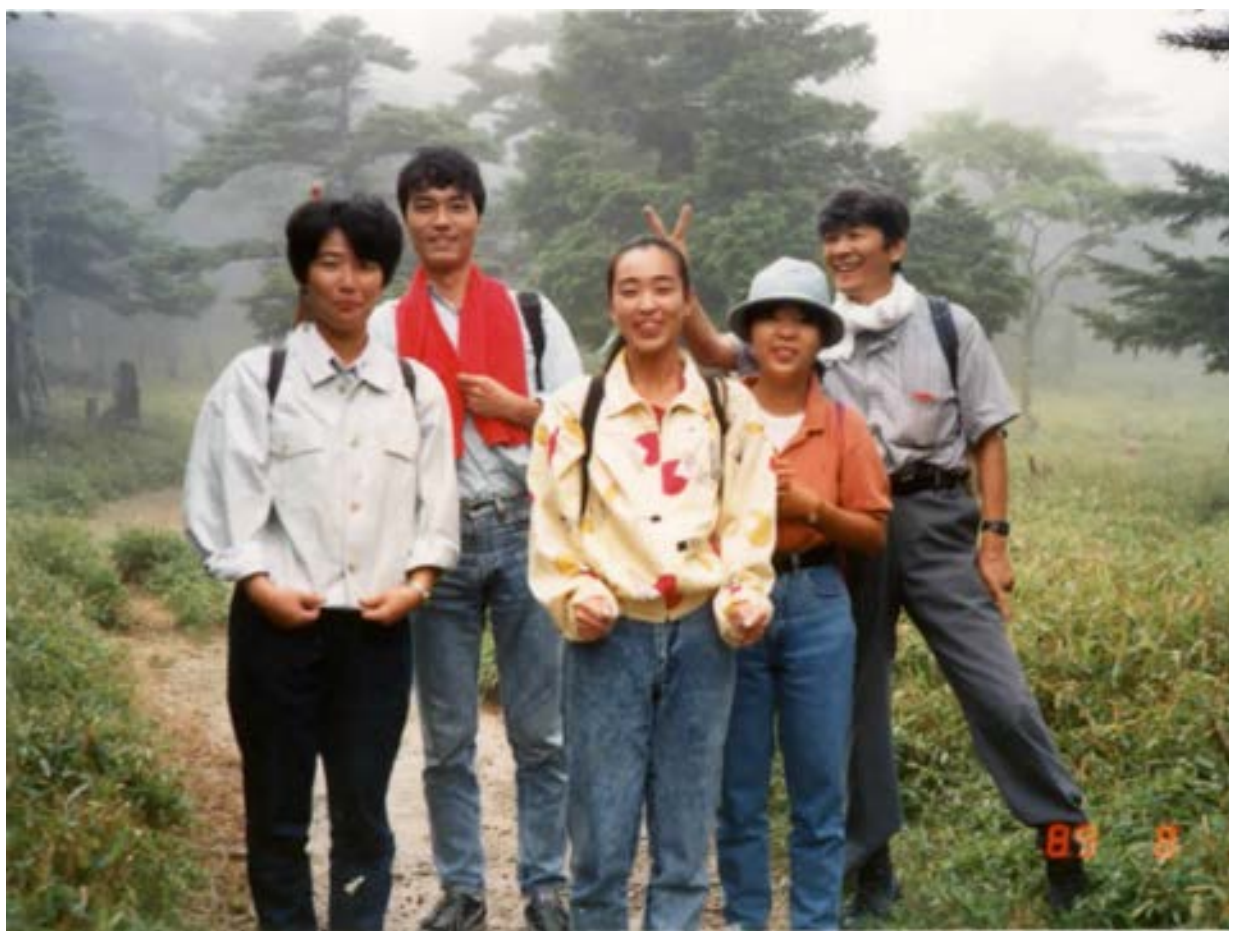

左から2番目が岡本、右が竹腰。
Okamoto 2nd from left, Takekoshi at right.

## Memories with Professor Takekoshi

Hiromi Okamoto

I joined the Takekoshi group in the spring of 1984. At that time, the laboratory was still in the Keage area. Old-style trams were running on Sanjo-dori right in front of the lab building. The gorgeous Miyako Hotel stood across the street, overlooking the ancient brick building in which we were working. The number of students were limited because the lab was far from the main campus of Kyoto University. I am the last graduate student who earned a Ph.D under the guidance of Prof. Takekoshi. He retired just several months after I got my degree.

For me, Takekoshi-sensei ("sensei" means "mentor" or "teacher" in Japanese) was more like a "father" than a "supervisor". My fond memories are mostly what happened when we drank, travelled, and had a quarrel, rather than scenes when we were working on physics. Other pupils of him will write about his physics achievements, so I would here like to share some trivial personal stories, calling his happy face to my mind.

There was the so-called "Tako-beya (octopus' room)" in the physics-department building on the main campus. All master's students belonging to the Division of Physics II were supposed to spend the first year there. The Takekoshi Lab was in Keage, far away from the Tako room, and furthermore, I used to be a lazy student playing around every day unlike other brilliant model fellows of the same age, .... so I rarely met my supervisor through the first fiscal year. One day near the end of the year, I received a telegram at my humble apartment, which made me quite surprised because it was my first time to receive a telegram. I first thought "perhaps my dad died....?", but the text was: "I have something to discuss with you right away, so please stop by at the laboratory. --- H. Takekoshi" As I was poor, I did not have a telephone and, in addition, I rarely visited Keage. The use of a telegram was the only way for Takekoshi-sensei to contact me. An average professor would probably think about cutting off such a terrible pupil unworthy of his supervisor. BUT, fortunately for me, Takekoshi-sensei was different. He warmly watched over me even though I was doing stupid things! By the way, I received a telegram from him one more time a year or two later. I do not remember exactly when, but at that time, I was sick in bed and was absent for more than a week without leave. I then got a telegram again which read "I haven't seen your face for a while in the lab, are you OK? --- H. Takekoshi". Well, .... I realized that I had made my supervisor get worried pretty much.

In the second year of the master's program, I came to show up in the Keage lab, I think, every day. A budget was approved to construct a new low-energy proton linear accelerator as a replacement for the old cyclotron. I, of course, joined the R&D team. The operating frequency was set at a very high value, i.e., 433.3 MHz, which required us to employ some cutting-edge technologies. My first research theme was about the design of a permanent magnet quadrupole (PMQ) lens. I spent a lot of time with Takekoshi-sensei to design and construct a unique system for measuring the magnetization direction of magnet pieces before they were assembled into a lens. I also made a compact rotating-coil device that quickly analyzes the multipole field within a PMQ aperture. Dr. Iwashita, who was an assistant professor of the Yanabu group, helped me to complete the electronic circuit system. Mr. Masaru Sawamura, the only graduate student besides me at that moment, was working on the post-coupler used to improve the stability of the electromagnetic field in the Alvarez-type drift-tube linac (DTL). The PMQ lenses I assembled then are still inside the drift tubes of the 7MeV DTL operational on the Uji campus of Kyoto University.

I finished my master thesis work relatively early and had time to spare, which gave me a chance to read a bunch of papers on radio-frequency quadrupole (RFQ) linacs I was personally interested in. The RFQ is now a standard ion injector, but at that time, no practical model existed in my country. There were no experts around me, so I studied the basics mainly through papers written by accelerator researchers in the United States. One day, Takekoshi-sensei came to my desk and said, "Okamoto-san, you know, we have to design an RFQ here. You are probably most familiar with that linac structure, so I want you to do that job." It is ridiculous to ask a fledgling graduate student to do such an important thing, but it is a decision that could be made only by Takekoshi-sensei. There was also somebody else (an assistant professor of the group) working on the RFQ design with the PARMTEQ code, but in the end, the parameters I proposed were adopted.

It happened in a regular group meeting. I reported on the result of my calculations regarding the RFQ's vane-tip modulation geometry or something. Takekoshi-sensei made a comment like "That does not really matter". I got a bit offended by his comment and replied "It should matter". I finished my presentation, leaving a parting shot: "OK, if you say that, why don't you do the rest of the calculations yourself?". After the meeting, a short while later, he came to me and said with a soft look, "Okamoto-san, you were angry at me this morning, weren't you? .... Would you like to go out for a drink together tonight?"

There was a tram stop right in front of the Keage Lab. We could go straight to the downtown Sanjo Kawaramachi very easily at any time. Takekoshi-sensei often took me to sake bars. Iwashita-san and Sawamura-san usually joined us. After having fun at an izakaya-type place, we visited sometimes a swanky bar close to the Keihan Sanjo station. During the bar-hopping, a poor student like me did not have to pay anything at all! I have never paid even a cent when I drank with Takekoshi-sensei.

When attending domestic accelerator-related meetings, we used to use a car to save travel expenses. Takekoshi-sensei gave young members a ride on his minivan and we ran on the highway in the middle of the night. If I remember correctly, the driver was mostly Sensei himself; at least I did not, because I had no license. When he became sleepy, he often ate a piece of chocolate. When getting even more tired, he starts to fold his hands on the steering wheel and put his chin on it. He turns the wheel maintaining such a dangerous driving posture. It was scary.... I could not distinguish whether he turned the wheel precisely along the curve or the movement of his niddle-noddle head happened to synchronize with the curve.

I continued the R&D for proton linear accelerators after I advanced to the doctoral course. I, however, wrote my Ph.D thesis on a theme that has nothing to do with it. While studying DTL, I got interested in alternating phase focusing (APF) and decided to explore its theoretical aspect. That was not a recommendation by somebody but I just started the work on my own. Since APF was a very minor type of accelerating structure, most of the available literature were written in Russian, impossible for me to understand. Takekoshi-sensei lent a helping hand to his student who had trouble with Russian. He brought a Russian-Japanese dictionary from somewhere and translated some papers for me. I can recall the image of my supervisor struggling with Russian papers in the lounge.

While I was working on my doctoral dissertation, Bob Jameson, one of Takekoshi-sensei's close friends, visited Japan on a sabbatical leave and stayed in Kyoto for six months. The friendship between Takekoshi-sensei and Bob continues uninterruptedly until his death. Bob gave me some APF papers (already translated in English!) and also many valuable comments on my dissertation. Getting to know Bob through Takekoshi-sensei had a big impact on my career as a physics researcher (of course, a very positive impact).

After getting Ph.D, I crossed the ocean to the US on the recommendation of Bob. I was originally determined to stay there for a long period but ended up leaving the country only in half a year because Takakoshi-sensei called me back to put me in an assistant professorship position of the lab. The return flight from Washington D.C. arrived at the airport in Osaka late at night. I found Sensei waiting for me at the arrival gate. On the way to Kyoto by his car, he said "you probably miss Japanese food". We stopped by at a small restaurant after midnight and he treated me to a Japanese meal which I enjoyed very much.

Two years later, I married in Kyoto. Needless to say, the matchmakers are Mr. and Mrs. Takekoshi. Bob also participated in the wedding ceremony. In the opening address from the matchmakers, Takekoshi-sensei formally introduced me to the participants. Instead of making compliments, he started to criticize me (of course, not seriously), stating "When Okamoto-san was a student, he came to the lab around 2:00 pm every day and disappeared at 6:00 pm already." At that time, I didn't really care about what he said, but now I would like to add one word here: "Sensei, 2 pm is an exaggeration. I was there at work at least around noon!"

Shortly after retirement, Takekoshi-sensei was invited by a private university in the Hiroshima area and filled a professorship position there. I once visited his apartment close to downtown Hiroshima. He liked gardening very much. He put some foliage plants in his room whose floor had got soiled from the dirt overflowing from the flowerpots. He showed me to his laboratory where he was playing with the Stirling engine he made himself. He also paid attention to solar cells, foreseeing future energy problems. I think that he had already bought some solar-cell samples to launch the detailed research. That night, we drank at a couple of bars in the downtown. We returned the apartment late, and I went to sleep between the plants.

In the fall of 1998, I moved to Hiroshima, leaving Kyoto where I spent many years since entering the university. Takekoshi-sensei was still working at the private university in Hiroshima. If I remember correctly, he retired the next spring and then returned to Kyotanabe where he owned a house. So, for at least the first several months, my family and I were in a place where it took less than an hour by car to his home. He sometimes paid a visit to us with freshly harvested seafood. We even went on a day trip together to some sightseeing spots and hot springs nearby. My wife remembers, better than me, when we drove to a kind of theme park where visitors can enjoy interacting with small animals. Takekoshi-sensei made the plan to please my elder son who was one year old. My wife says, "You walked away alone, leaving me behind". She was pushing the stroller, even with some heavy baggage. I heard later that Takekoshi-sensei kept accompanying my wife, carrying the whole baggage for her bum husband. By the way, at this time, he should have completed the scale model of a solar car already at his lab.

After Takekoshi-sensei returned to Kyotanabe, I had almost lost the opportunity to meet him. He, however, suddenly mailed us a package of regional specialties sometimes. I had regularly made a phone call to him around O-Bon and also near the end of every year to know how he is, but finally lost contact a few years ago.

I can't remember when it was, .... I met Takekoshi-sensei some years later and entered a sake bar near the Kyoto station when I visited there on a business trip or something. I made a phone call to him to freeload as usual. That was the last time I saw him. While I can't remember how many years ago it was, I can clearly remember the seats, the atmosphere of the bar, some of the dishes I ordered, and above all, the Sensei's smile.

Hiromi Okamoto

# Personal Memories – Bob Jameson

### In Memory of my friend Prof. Hidekuni Takekoshi

In Kyoto at the Keage Institute of the Institute for Chemical Research, Kyoto University, I first met Prof. Hidekuni Takekoshi. I will abbreviate his name as "T". He was very interested in PIGMI, also enjoyed hosting foreign visitors, and introduced many things even on this first meeting. He visited Los Alamos in 1981, and also sent Y. Iwashita, his young staff member, for three years to work with us in AT Division. It was already clear that we were to be friends, and for my 1981 return visit to Japan, for which I requested that he arrange a Japanese-style accommodation, he replied that I would stay "at his house". Remarkable, as that would be very unusual anywhere, but perhaps especially in Japan. So I traveled without accommodation reservation.

"His house" was two houses – one in Tokyo, one in Kyoto. His father had built two houses in a small walled garden in north-central Tokyo. The second had been for use of the four brothers at various times, but now mostly empty and available to T when he was in Tokyo. Later I had my own key to this house for many years – that was quite something – to be able to go to Tokyo at any time and stay at "my own house"! In Kyoto, T had cleverly invested in a small piece of land in a development planned to be near a new "Kansai Science City" corresponding to the northern Kanto area Tsukuba, where KEK is located. We also started "going around" to beautiful tourist areas of Japan.

We were then, and remained, best friends. His friendship has been a major influence on my life – a true "companion along the way".

Since 1980, I have averaged one trip per year to Japan, mostly for at least one month, often for several months, once for a year, and explored with T almost every corner of the country. Words were not so many, although his English was quite ok – in Japan it is necessary to be able to learn through feelings. It is necessary to at least learn the two supplementary phonetic tables – katakana and hiragana – they are used for almost all foreign words and it helps very much to know it. The Chinese characters, reading, real Japanese conversation would have required too much time then, with so much to do, computing, etc. So one could not be pestering continuously with requests for translation. And he indicated that very soon, when I asked "What does "chotto-mate" mean?" after hearing it all the time. He replied "You will figure it out…" So I grasped completely that feeling was the key, and figured out eventually that "chotto-mate" meant "Wait a moment…"

After my decision to do something else at the end of 1987, I informed T, without making any requests. He immediately went to work, and although the Japanese fiscal year was already half over, he managed an invitation from KEK for a year's visit. I informed KEK that I was very glad for their invitation, but that it was necessary for me to spend at least half of the year with T in Kyoto. They had no objection at all. The old Keage Institute had become a very special place for me. It had been built in the Meiji Era, when Japan was opened to the West, as the first hydroelectric power plant in Japan, at the end of a long tunnel/canal from Biwa Lake (Biwako), in the Meiji red-brick style, with very thick walls.

After WWII and the infamous incident, when the occupation forces destroyed the original Kyoto cyclotron and all the other Japanese cyclotrons to prevent further nuclear researches, the Keage hydroelectric plant was replaced by a new one next door, and the old building was ideal for the building of a new cyclotron. T was a student and fortunate to be in Kyoto during the war, and helped finish the cyclotron, which became one of the longest running cyclotrons in the world. He then went to the Japan Atomic Energy Research Institute (JAERI) in northern Japan, and extended the electron linac from 20 MeV to 120 MeV. From there he returned as Professor of the Keage Institute. Working in the Keage Building was so peaceful and wonderful, it was the feeling of being a monk in a temple. I was determined to spend a longer period there, and 1988 was the last chance, as a new building on the KU ICR Uji campus outside Kyoto would be ready in 1989. Among many, it was a major reason why I changed course in 1987 – I would have set aside time in 1988 in any case and under any circumstance.

During the 1988-1989 year, long linac investigations continued and also work with then-student Hiromi Okamoto on APF schemes, which he expanded and wrote a paper which I and others found very useful, even many years later.

A further activity for 1988-1989 was to become familiar with the program for atomic waste transmutation at JAERI (later JAEA). The Japanese had started a development program, and I arranged to meet them and learn about it.

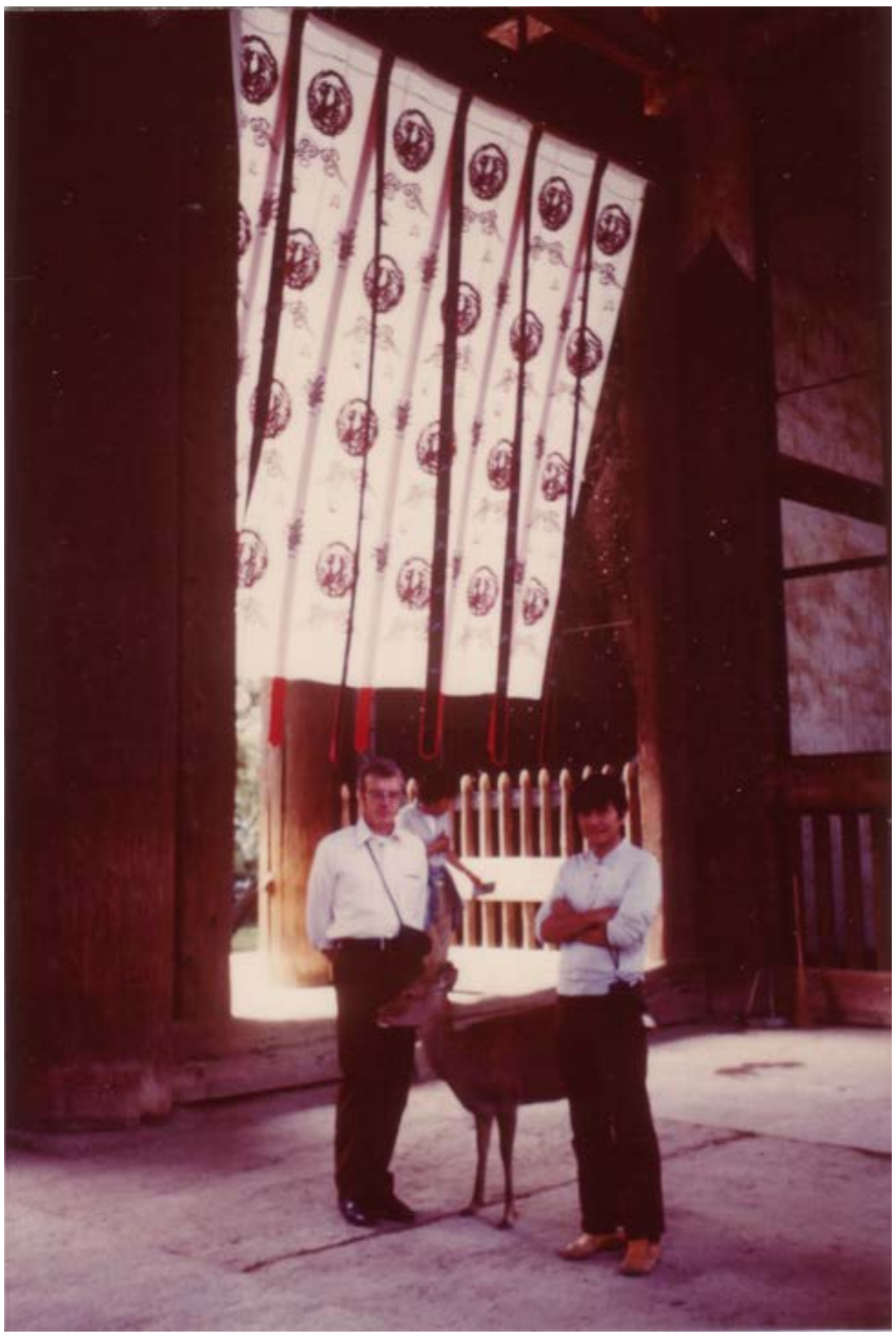

1981

At some point, T told me that I had a new job. JAERI was founding a new accelerator group for ADT (Accelerator-Driven-Transmutation), the Group Leader Mizumoto had been a member of T's linac group at JAERI, a nuclear physicist using the electron linac but overnight ordered to become an proton accelerator expert. And I was to be his mentor. This was a very good job. JAERI contracted with me through LANL and provided funds that supported me and also were distributed by me in the LANL group to those who helped produce a succession of lengthy reports on all aspects of linac technology, and especially on my work on linac design. The Japanese grasped the advantages of a beam equilibrium, and the resulting linac was the first, and more or less still the only, long linac in the world with a fully equipartitioned design. It also has fully adjustable quadrupoles. The ADT project was merged with a long-sought project that Hirao had wanted for INS in Tokyo, later merged with KEK, for a spallation neutron source, and the J-Parc project was located at Tokai on the JAEA (JAERI) site. The ADT goal became low priority (and still awaits completion of the linac to ≥600 MeV), but the linac design remained, and it operates well as the J-Parc injector.

From ~1990 to 2017, many weeks were spent at Tokai-mura. On weekends during those many years, I would go to Tokyo for two days, and stay at Takekoshi's house, in the beautiful small compound in Tabata. It was a great honor – I could escape Tokai, where there was little to do, go Scottish dancing on Friday (and occasionally Saturday) evenings, and visit flea and antique markets on Saturday and Sunday, searching for mingei omiyage.

Explorations of Japan with T really began in earnest during the 1988-1989 year. After that, every visit included, as first priority, time with him. He retired from Kyoto University in March 1990, and went to work in Hiroshima, where I visited him every year, and then he moved back to Kyoto. I was able to visit him almost every year up to April/May 2019. The many adventures are related in my travel logs and diaries, and documented with many photos. The comments about driving are echoed – always with one hand. Driving in reverse was not part of his repertoire. If missing a turn on a major road and wanting to reach an opposite corner, always use small roads and "There will be a way". Once that lead to smaller and smaller roads and finally into a path in an orchard, with nowhere to turn around. I had to drive in reverse, up hill around curves, thru a small group of houses, for about one kilometer - the only casualty was one flower pot. We always made it, and were happy. I will miss him a lot.

His resting place is Kaizenji Temple, 3 Chome-3-3 Matsugaya, Taito City, Tokyo, Japan 111-0036. The closest station to Kaizenji is Asakusa Station.
<https://goo.gl/maps/9uekpabtkFsqhGNK7

R. A. Jameson (Retired)

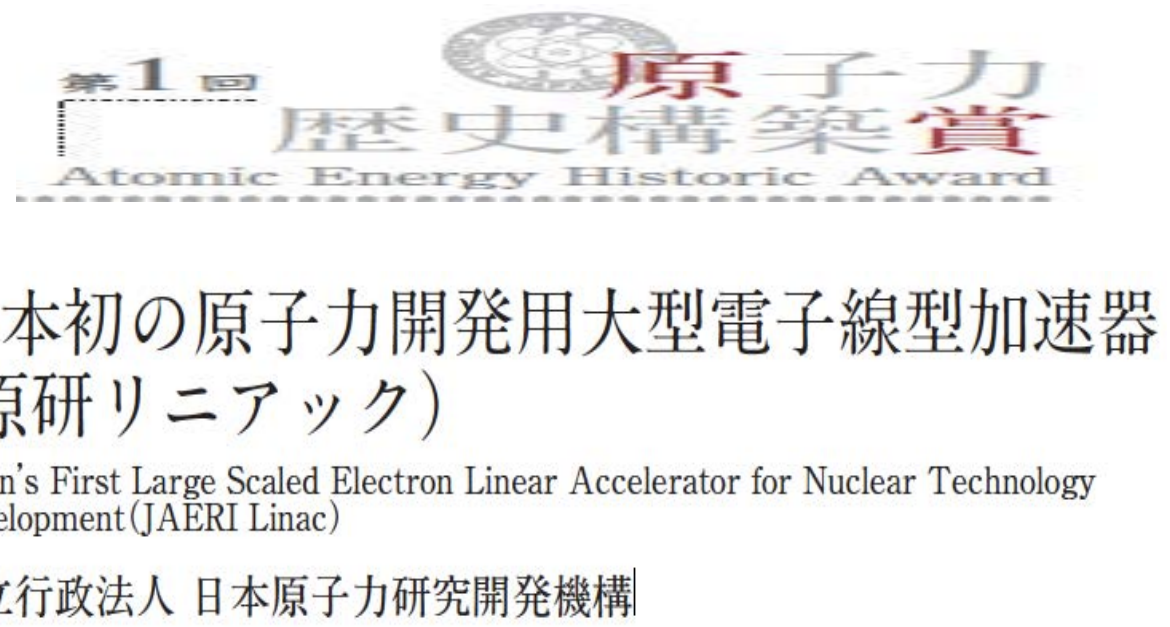

Keage Laboratory of Nuclear Science Decennial Report 1978-1988 (Commemoration Issue Dedicated to Professor Hidekuni Takekoshi On the Occasion of His Retirement), Takekoshi, Hidekuni, Bulletin of the Institute for Chemical Research, Kyoto University (1990), 68(2): 71-85, 1990-10-31, http://hdl.handle.net/2433/77340

Early in Taipei Imperial University and Kyoto University Accelerator development and nuclear physics research (Part 1), Takekuni Hidekuni,

(Early days of accelerator development and nuclear physics, Experiments at Daihoku Imperial University and Kyoto University (1/2), Hidekuni Takekoshi), J. Particle Accelerator Society, Vol 3, 4, 2006 (384-3, 46